\documentclass[journal=jctcce,manuscript=article]{achemso}
\usepackage{amsmath}
\usepackage{graphicx}
\usepackage{booktabs}
\usepackage{threeparttable}
\PassOptionsToPackage{hyphens}{url}
\usepackage{hyperref}
\usepackage{xcolor}

\makeatletter
\g@addto@macro{\UrlBreaks}{\UrlOrds}
\makeatother

\graphicspath{{figures/}}

\title{Compression of virtual spaces in transcorrelated methods via singular value decomposition: application to the G2 set}
\author{Yifan Cheng}
\altaffiliation{Y.C. and K.S. contributed equally to this work.}
\email{y.cheng@fkf.mpg.de}
\author{Kristoffer Simula}
\email{k.simula@fkf.mpg.de}
\altaffiliation{Y.C. and K.S. contributed equally to this work.}
\author{Johannes Hauskrecht}
\author{Evelin Martine Corvid Christlmaier}
\author{Daniel Kats}
\author{Ali Alavi}
\affiliation{Max Planck Institute for Solid State Research, Heisenbergstra\ss e 1, 70569 Stuttgart, Germany}
\alsoaffiliation{Yusuf Hamied Department of Chemistry, University of Cambridge, Lensfield Road, Cambridge CB2 1EW, UK}
\email{a.alavi@fkf.mpg.de}

\begin{document}
\maketitle

\begin{tocentry}
\includegraphics[width=\textwidth]{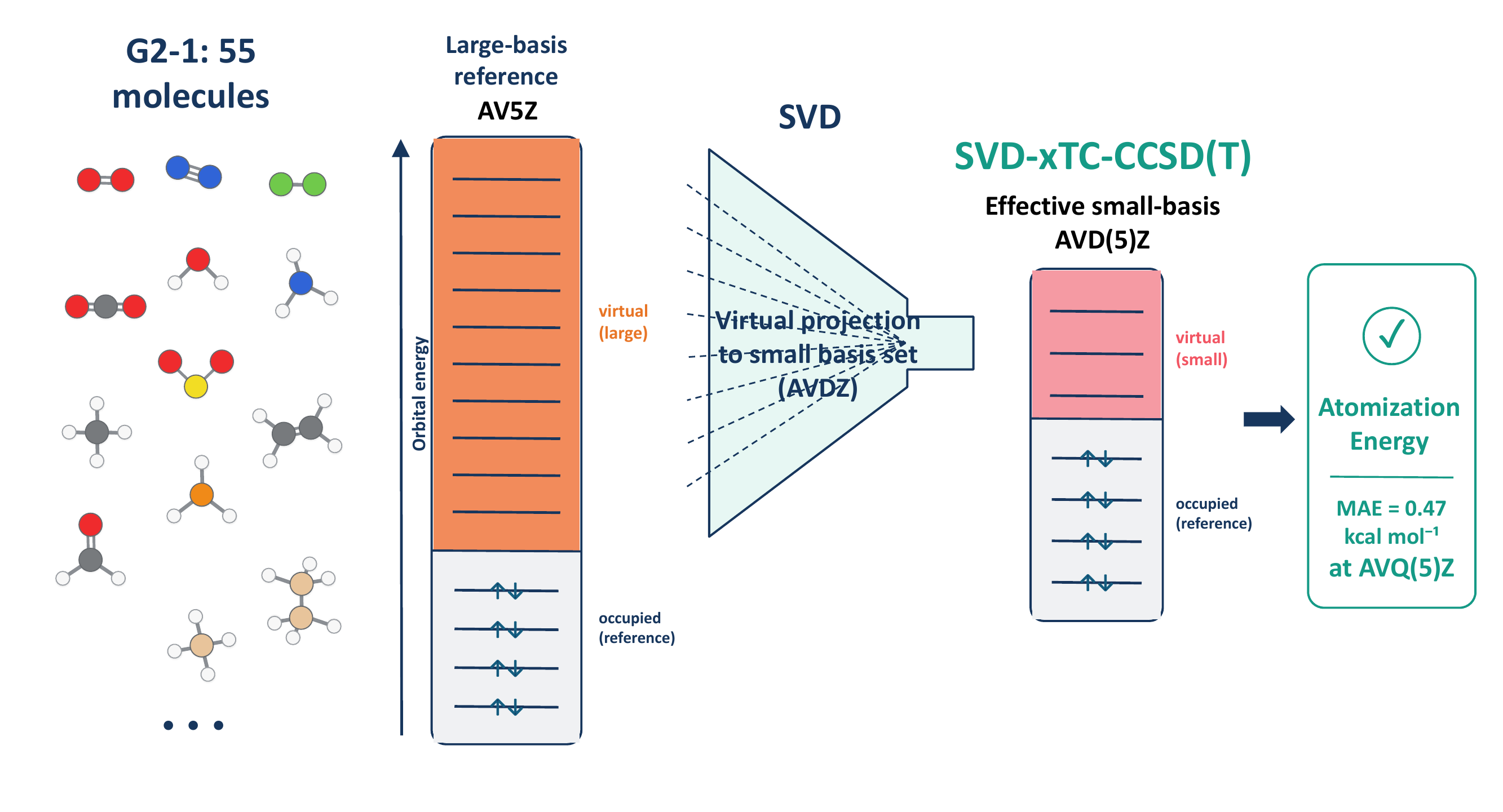}
\end{tocentry}

\begin{abstract}
We introduce a new singular-value-decomposition-based scheme for constructing small virtual spaces out of large basis sets for transcorrelated (TC) calculations, termed SVD-xTC. This work builds on the recent finding that the residual basis error in the TC reference energy converges more slowly than that of the correlation energy. Within the new workflow, the post Hartree-Fock TC calculation is performed in a compressed virtual orbital subspace, obtained by projecting the canonical virtual orbitals from a large basis set onto a smaller basis set through singular value decomposition (SVD). This allows us to achieve the high accuracy allowed by the large basis, whilst the bottleneck steps - TC integral calculation and post-HF correlation method such as CCSD(T) - incur the cost of only a small virtual space calculation. The method therefore is highly efficient, whilst avoiding the composite nature of the reference correction method. Using the new scheme, we widen the scope of benchmark-quality TC results into more complex molecules than previously considered: using the G2-1 set of 55 molecules with first- and second-row atoms, we apply SVD-xTC-CCSD(T) to compute atomization energies. We compare our results against the near-exact semistochastic heat-bath configuration interaction (SHCI) reference values and experiment. We find that SVD-xTC-CCSD(T) delivers chemical accuracy already with triple-$\zeta$ basis sets. Finally, we use the quadruple-$\zeta$ results to analyze the accuracy of pseudopotentials within the TC method, and show that pseudopotential TC workflow provides faster basis-set convergence than all-electron TC. We also present timings for computing the atomization energies on G2-1 set, demonstrating the efficiency of our TC workflows. 

\end{abstract}

\section{Introduction}
In electronic structure theory, a calculated value is often considered quantitatively, or "chemically", accurate when it agrees with a chosen reference value to within $1~\mathrm{kcal\cdot mol^{-1}}$ ~\cite{karton2022}. This energy scale corresponds to only 1\% of a typical covalent-bond dissociation energy, and approximately twice the value of $k_{\mathrm B}T$ at room temperature, placing it on an energy scale relevant to thermal fluctuations. Although finer scales are sometimes relevant in contemporary chemistry, for example in describing weak interactions or calculating vibrational spectra, chemical accuracy remains a standard benchmark in electronic structure theory. Achieving it requires both a sufficiently complete treatment of electron correlation and a sufficiently complete basis set -- a combination that, given the rapid growth of the computational cost of correlation methods with system size, generally remains practical only for small to intermediate molecules. Even for these, the computational cost of reaching chemical accuracy can be prohibitive. Thus the need for methods that reach chemical accuracy for a wide range of systems with modest computational effort  remains a central challenge in electronic structure theory.

Transcorrelation (TC)~\cite{boys1969determination,boys1969condition,boys1969calculation,ten2000feasible,umezawa2003transcorrelated,umezawa2005practical} addresses this goal by modifying the Hamiltonian itself rather than the basis or the correlation solver: a similarity transformation with a real-space Jastrow correlator explicitly builds in the interparticle cusps, which wavefunction expansions constructed purely from antisymmetrised single-particle products cannot capture without significant effort. The similarity transformation preserves the eigenspectrum of the original Hamiltonian while drastically accelerating basis-set convergence and pulling correlation from the virtual into the occupied space, compactifying the wave function so that lower-level second-quantized correlation solvers already yield converged energies. 


In recent years, TC theories have been improved on many fronts: the problematic $3$-body terms can be treated very accurately with cheaper $2$-body corrections via the xTC approximation~\cite{christlmaier2023}, highly accurate schemes for both VMC-based~\cite{haupt2023} and deterministic~\cite{filip2025jastrow} optimization of Jastrow factors have been established, and the route to larger systems with heavier atoms has been opened by the establishment of the TC-pseudopotential theory~\cite{simula2025ecp}. Many theoretical and methodological extensions and applications of TC have also appeared, including the transcorrelated coupled-cluster (TC-CC) framework~\cite{schraivogel2021,schraivogel2023,kats2024}, TC full configuration interaction quantum monte carlo (TC-FCIQMC)~\cite{luo2018combining,haupt2025}, TC density matrix renormalization group (TC-DMRG)~\cite{baiardi2022explicitly,liao2023dmrg,corbett2025scaling}, TC selected configuration interaction (TC-sCI)~\cite{ammar2024compactification}, F12-inspired transcorrelation methods\cite{yanai2012canonical,ten2023nonunitary}, modular jastrow construction~\cite{haupt2026modular}, quantum computing~\cite{sokolov2023orders,dobrautz2024toward,li2024variational,Uvarov2026}, accurate calculations on second-row elements~\cite{filip2025}, transition-metal atoms~\cite{simula2025tc_transition_metals}, and solids~\cite{liao2021,simula2026silicon}.

Recently, it has been numerically shown (for the HEAT dataset of molecules) that the accelerated basis-set convergence of the TC total energy arises primarily from the TC correlation energy, whereas the basis-set convergence of the TC reference energy is not significantly improved.~\cite{hauskrecht2026} This means that, in contrast to conventional wave-function calculations, in which the correlation energy converges more slowly than the reference energy with increasing basis-set size, the TC correlation energy converges faster than the TC reference energy. Consequently, the error in the reference energy becomes the dominant bottleneck in basis-set convergence. The suggested \textit{reference correction} (RC) scheme can correct the total TC energy in a small basis by using a reference energy from a large basis. This leads to significant improvements in the accuracy of TC methods, especially in small basis sets. 
However, RC-TC works in a composite scheme, which requires two reference energy calculations and thus unnecessarily complicated computational workflow.

Here we develop and apply a more practical scheme that retains an accurate reference orbital set throughout the TC workflow, but compresses the virtual space to the size of a smaller basis. We refer to this SVD-based scheme as SVD-xTC. SVD-xTC no longer follows a composite-energy-like formulation and provides a simple one-step TC workflow. It also has an improved occupied space compared to standard small-basis workflow due to accurate reference orbitals. In the following text, we use xTC abbreviation to refer to our TC calculations (since all calculations employ xTC approximation), but use TC to refer to the general transcorrelated framework.

We apply SVD-xTC to the G2-1 set,~\cite{curtiss1991} a standard benchmark comprising 55 small first- and second-row molecules with available experimental atomization energies. Method-to-method comparisons are made against near-exact SHCI atomization energies extrapolated to the complete-basis-set limit ~\cite{Yao2020}. By comparing the results with non-TC and standard xTC coupled cluster, We find that SVD-xTC delivers the most accurate CCSD(T) results. We also find the SVD-xTC results to be chemically accurate in atomization energies against both SHCI+PBE+CV and experiment, even when CCSD(T) is used as the correlation solver. Finally, we benchmark the pseudopotential TC calculations with both He- and Ne-core ccECPs~\cite{bennett2017,annaberdiyev2021}, and quantify the inherent ECP error. The Ne-core ccECPs are found to converge faster with respect to basis size, but have larger error against all-electron results, while He-core ccECPs converge similarly to all-electron calculations, with very similar accuracy. Together, these calculations constitute an extensive benchmark of TC methods for the challenging prediction of atomization energies and provide a basis for discussing future developments of TC theory.

\section{Theory}

In this section, $p,q,r,s,\ldots$ denote general orbitals, $i,j,\ldots$ denote occupied orbitals, and $a,b,\ldots$ denote virtual orbitals.  For simplicity, we employ the spin-orbital formalism.


The similarity transformation, expressed as a commutator expansion via the Baker-Campbell-Hausdorff formula, yields the transcorrelated Hamiltonian $\hat H^{\text{TC}}$~\cite{cohen2019}:
\begin{align}
    \begin{aligned}
        \label{eq:simtransf}
\hat H^{\text{TC}}
&= e^{-J_\alpha}\hat H e^{J_\alpha} = \hat{H}+\left[\hat{H},J_\alpha\right]+\frac{1}{2}\left[\left[\hat{H},J_\alpha\right],J_\alpha\right], 
\end{aligned}
\end{align}
where $J_\alpha$ is the Jastrow factor, and $\alpha$ represents a set of parameters within the Jastrow function, optimized using a TC-adapted form of variational Monte Carlo (VMC) \cite{haupt2023}. In the commutator expansion, only terms of $\hat{H}$ that do not commute with $J_\alpha$ survive and give contributions to the TC hamiltonian: the kinetic energy operator, and, if used, the pseudopotential operator. 
We call the $2$- and $3$-body terms arising from the kinetic energy commutators $\hat{K}_\alpha$ and $\hat{L}_\alpha$, respectively, so that $\hat K_\alpha+\hat L_\alpha=\left[\frac{1}{2}\sum_p\nabla_p^2,J_\alpha\right]+\left[\left[\frac{1}{2}\sum_p\nabla_p^2,J_\alpha\right],J_\alpha\right]$, and the pseudopotential commutators we define as $\hat P_\alpha=\left[\sum_p\hat{V}_{ecp}^p,J_\alpha\right]+\left[\left[\sum_p\hat{V}_{ecp}^p,J_\alpha\right],J_\alpha\right]$. In the all-electron case (no pseudopotentials), the commutator expansion terminates after the second commutator. The resulting $\hat{H}^{\text{TC}}$ contains up to $3$-body terms, which are approximately treated as $2$-body terms in the xTC approximation~\cite{christlmaier2023}. When pseudopotentials are used, the expansion is not guaranteed to terminate after the second commutator; in earlier work, we have shown that only the first two commutators need to be evaluated, and we make the further approximation of ignoring the $3$-body terms arising from the pseudopotential operator, under the assumption that the contribution of three correlating electrons in the ECP core region is small~\cite{simula2025ecp}.

In second quantized basis, the TC Hamiltonian with the xTC approximation, which we employ throughout this work, takes the following (non-hermitian) form:
\begin{align}
    \begin{aligned}
H^{\text{xTC}}&=E^{\text{xTC}}_0+\sum_{pq} {h}^{\text{xTC}}_{pq}\, a_p^\dagger a_q
+\tfrac12 \sum_{pqrs} {V}^{\text{xTC}}_{pqrs}\, a_p^\dagger a_q^\dagger a_s a_r
\end{aligned}
\end{align}
The operators $a_p^\dagger$ and $a_p$ are the creation and annihilation operators of the orbitals $\phi_p(\mathbf{r})$. The two-body xTC interaction term ${V}^{\text{xTC}}$ is defined as
\begin{align}
    \begin{aligned}
    \label{eq:T_pqrs}
{V}^{\text{xTC}}_{pqrs}
&= {V}_{pqrs} + \Delta V_{pqrs}, \\
\Delta V_{pqrs}&=  - {K}_{pqrs} + {P}_{pqrs}
+\Delta {W}_{pqrs}, \\
\Delta {W}_{pqrs}&= - \sum_{tu}\left({L}_{prtqsu}
                                        -{L}_{prtqus}
                                        - {L}_{prtusq}\right)\gamma_{tu}.
    \end{aligned}
\end{align}
Above, $\Delta V_{pqrs}$ is the xTC correction to the two-body term, consisting of $2$-body contributions arising from the kinetic energy operator ($K_{pqrs}$), the pseudopotential operator ($P_{pqrs}$), and the xTC approximation ($\Delta W_{pqrs}$). $\gamma_{tu}$ is the one-body reduced density matrix of the reference state.

The one-body and constant xTC terms are defined as:
\begin{align}
    \label{eq:F_pq}
    \begin{aligned}
      {h}^{\text{xTC}}_{pq} &= {h}_{pq} + \Delta {h}_{pq},\\
      E^{\text{xTC}}_0 &= E_0 + \Delta E_0^{\text{xTC}}.
    \end{aligned}
  \end{align}
 where the correction defined under a single determinant reference approximation can be simplified to~\cite{christlmaier2023}:
  \begin{align}
    \label{eq:e0-xtc}
    \begin{aligned}
      \Delta {h}_{pq}&= -\frac{1}{2}\sum_{tu}\left(\Delta {W}_{ptqu}-\Delta {W}_{ptuq}\right)\gamma_{tu}\\
    \Delta E_0^{\text{xTC}}&=-\frac{1}{3}\sum_{tu}\Delta h_{tu}\gamma_{tu}.
\end{aligned}
\end{align}

We employ a Drummond-Towler-Needs-type Jastrow factor~\cite{drummond2004} with two-body ($u$), one-body ($\chi$), and three-body ($f$) terms in this work:
\begin{align}
  J_{\alpha_u,\alpha_\chi,\alpha_f}=\sum_{i\neq j}u_{\alpha_u}(\mathbf{r}_i,\mathbf{r}_j) + \sum_{I,i}\chi_{\alpha_\chi}(\mathbf{r}_i,\mathbf{R}_I) + \sum_{I}\sum_{i\neq j}f_{\alpha_f}(\mathbf{r}_i,\mathbf{R}_I,\mathbf{r}_j),
\end{align}
where $\mathbf{r}_i$ are the electron positions, $\mathbf{R}_I$ the nuclear positions, and $\alpha_u,\alpha_\chi,\alpha_f$ are the variational parameters. Each term is expressed as polynomial function and truncated at its own cutoff length $L_u$, $L_\chi$, $L_f$. The $u$ term enforces the electron-electron cusp condition. To improve the stability of the VMC optimization, the electron-nucleus cusp condition is not imposed through the $\chi$ term. Instead, an additional cusp-correcting Jastrow term, $\Lambda(r)$, is introduced for all-electron calculations.\cite{ma2005scheme,haupt2023} With ECPs it is not necessary to treat the electron-nucleus cusp. 

The xTC Hartree--Fock (HF) reference energy 
\begin{equation}
E^{\text{xTC}}_{\text{ref}} = \langle\Phi_0|\hat H^{\text{xTC}}|\Phi_0\rangle = E_0^{\text{xTC}} + \sum_{i} h_{ii}^{\text{xTC}} + \frac{1}{2}\sum_{ik} (V_{iikk}^{\text{xTC}} - V_{ikki}^{\text{xTC}})
\end{equation}
is not invariant with respect to the Jastrow factor. A properly optimized Jastrow factor captures the electron-electron and electron-nucleus correlations of the reference and thus lowers the reference energy. Since the similarity transformation preserves the exact eigenvalues of a second-quantized Hamiltonian in basis set limit, the lowering of the xTC-HF reference energy can be seen as transfer of correlation from virtual excitations into the occupied/reference sector. Thus, the remaining correlation energy to be captured by a second-quantized correlation solver is reduced, thereby reflecting the wave-function compactification achieved by the TC method.

Besides compactification, the Jastrow factor contains the cusp corrections and is optimized in continuum via VMC, thus accelerating the basis set convergence in the description of correlations via virtual excitations. However, the basis error of the non-TC reference energy, which is still part of $E^{\text{xTC}}_{\text{ref}}$, is not reduced~\cite{hauskrecht2026} , which motivates both the RC-xTC scheme of Ref.~\cite{hauskrecht2026} and the alternative SVD-xTC method developed here.

\begin{figure*}[htbp]
  \centering
  \includegraphics[width=\textwidth]{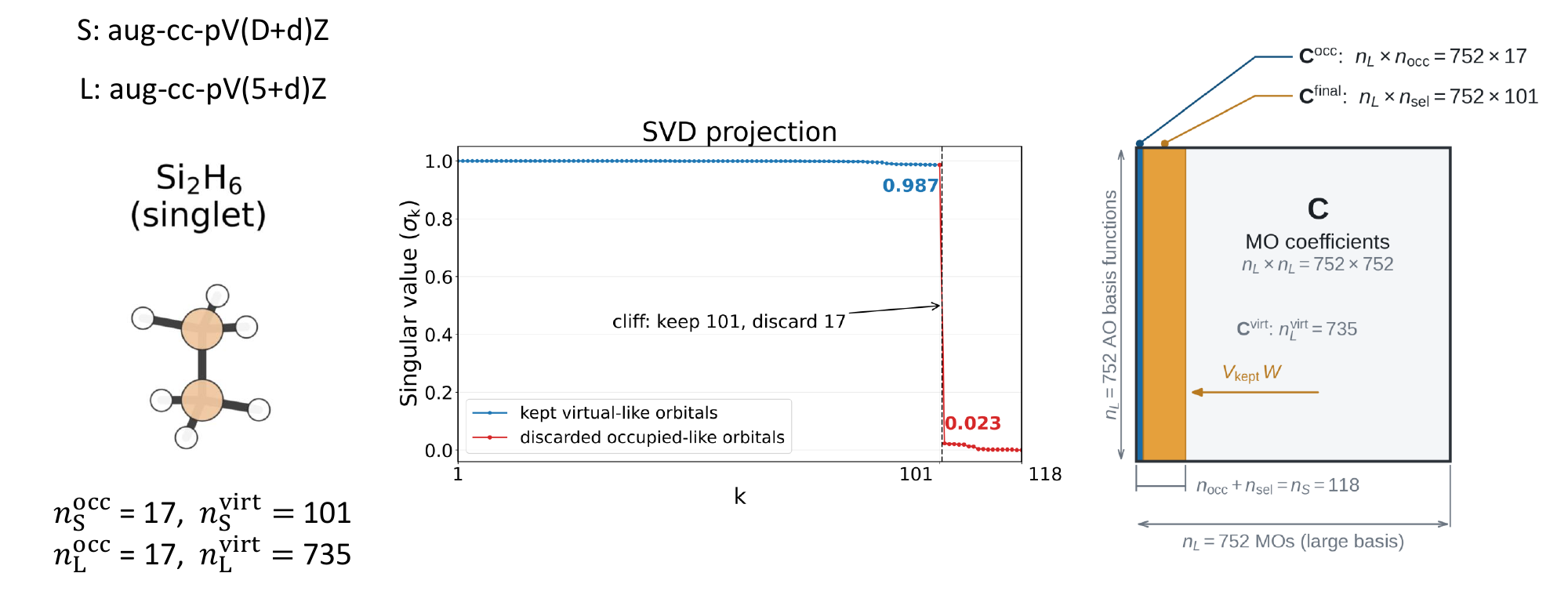}
  \caption{\label{fig:svd_example}Singular values $\sigma_k$ of the projection matrix $M$ for Si$_2$H$_6$, projecting the aug-cc-pV(5+d)Z virtual MOs onto the aug-cc-pV(D+d)Z AO space. The sharp cliff between $\sigma_{101}=0.987$ and $\sigma_{102}=0.023$ provides an unambiguous criterion for the choice of the projected virtual space. The rightmost panel illustrates the resulting truncation of the MO coefficient matrix: the retained orbital space has $n_{\text{occ}}+n_{\text{sel}}=n_\text{S}=118$ columns, matching the dimension of the small basis, and the remaining columns (gray) are discarded.}
\end{figure*}

In SVD-xTC, the selected virtual space is constructed by projecting the large-basis virtual molecular orbitals (MOs) onto the small-basis atomic orbital (AO) space via
\begin{equation}
  M = S_{SS}^{-1/2} S_{SL} C^{\text{virt}}
  \qquad [n_S \times n_L^{\text{virt}}]
\end{equation}
where $S_{SS}$ is the small-basis AO overlap matrix, $S_{SL}$ is the AO cross-overlap between small-basis $S$ and large-basis $L$, and $C^{\text{virt}}$ are the large-basis virtual MO coefficients. Here $n_S=\dim\mathcal H_S$ and $n_L=\dim\mathcal H_L$ are the numbers of small- and large-basis AOs. If $n_{\text{occ}}$ is the number of occupied orbitals in $|\Phi_{\mathrm{HF}}^{(L)}\rangle$, and the large-basis and compressed virtual counts are $n_L^{\text{virt}}=n_L-n_{\text{occ}}$ and $n_{\text{sel}}=n_S-n_{\text{occ}}$, respectively, the resulting compressed orbital space contains $n_{\text{occ}}+n_{\text{sel}}=n_S$ orbitals, matching the small-basis dimension.
The elements $M_{\mu a} = \langle\tilde\varphi_\mu|\psi_a\rangle$ are overlaps
between small-basis L\"{o}wdin-orthonormalized AOs (OAOs) and
large-basis canonical virtual orbitals, so the SVD
\begin{equation}
  M = U\Sigma V^{\top}
  \qquad
  U\ [n_S\times n_S],\;
  \Sigma\ [n_S\times n_S],\;
  V\ [n_L^{\text{virt}}\times n_S]
\end{equation}
yields singular values $\sigma_k\leq 1$ that are exact cosines of the principal
angles between the two subspaces. In $\Sigma$, the last $n_{\text{occ}}$ cosines should be near-zero: the large-basis virtual MOs are orthogonal to the large-basis occupied space, and therefore cannot span the full small-basis AO space. We find this behavior throughout the G2-1 dataset. The SVD projection of Si$2$H$6$, shown as an example in Fig.~\ref{fig:svd_example}, exhibits a sharply defined boundary between the two groups of singular values, allowing an unambiguous choice of the projected virtual space. We omit the right singular vectors corresponding to these $n_{\text{occ}}$ cosines, and construct a compressed virtual space by rotating the large-basis virtual MOs with the $n_{\text{selected}}=n_S-n_{\text{occ}}$ columns of $V$ collected in $V_{\text{kept}}$,
\begin{equation}
  C^{\text{selected}} = C^{\text{virt}} V_{\text{kept}}
  \qquad [n_L\times n_{\text{sel}}],
\end{equation}
which is $S_{LL}$-orthonormal by construction. The SVD rotation mixes
canonical virtuals, rendering the projected Fock matrix
\begin{equation}
  F^{\text{selected}} = V_{\text{kept}}^{\top}\,\mathrm{diag}(\varepsilon_a)\,V_{\text{kept}}
  \qquad [n_{\text{sel}}\times n_{\text{sel}}]
\end{equation}
non-diagonal. Diagonalizing $F^{\text{selected}} = W\tilde\Lambda W^{\top}$ yields
the compressed set of pseudocanonical virtual orbitals:
\begin{equation}
  C^{\text{final}} = C^{\text{virt}}\,V_{\text{kept}}\,W
  \qquad [n_L\times n_{\text{sel}}].
\end{equation}
The pseudocanonicalization of the compressed virtual orbitals is not strictly necessary, but we do it to accelerate convergence in subsequent SVD-xTC-CCSD(T) calculations. 
ElemCo.jl~\cite{elemcojl} includes the full non-Hermitian xTC Fock matrix, $f^{\text{xTC}}_{pq} = h^{\text{xTC}}_{pq} + \sum_{k}(V^{\text{xTC}}_{pqkk} - V^{\text{xTC}}_{pkkq})$, in the CCSD and $\Lambda$ equations, and uses its diagonal elements only as denominators in the iterative amplitude updates. The perturbative triples are evaluated in the $\Lambda$CCSD(T) formulation, after the occupied--occupied and virtual--virtual blocks of $f^{\text{xTC}}$ have been diagonalized by a biorthogonal transformation. The final energy is therefore invariant with respect to rotations within the compressed virtual space. The pseudocanonicalization nevertheless speeds up convergence. In the pseudocanonical basis, the conventional Fock matrix is diagonal. Since $f^{\text{xTC}}$ differs from conventional Fock only by the xTC corrections, $f^{\text{xTC}}$ is nearly diagonal as well. Its diagonal elements are then good denominators, and fewer CCSD and $\Lambda$ iterations are needed.

The resulting orbital space combines a reference taken from the SCF solution in the large basis with virtual orbitals expanded in the large-basis AOs but reduced in number to match the small basis. For open-shell systems, any orbital degenerate with an occupied orbital is added to the reference space before compression. The construction generalizes naturally to complete-active-space SCF (CASSCF) references: the inactive-plus-active space replaces the occupied-plus-degenerate set, and SVD compresses the remaining virtual space. 

In addition to leaving the occupied orbitals and reference determinant unchanged, SVD compresses the virtual space in a manner that preserves occupied-virtual orthogonality throughout the calculation, resulting in a well defined virtual space distinct from that of the original target basis. Therefore, the obtained energies are not expected to coincide with RC-xTC results, but are expected to be more accurate than the latter. This will be discussed in detail in the results section. 


\section{Computational Details}

\begin{figure*}[t]
    \centering
    \includegraphics[scale=.87]{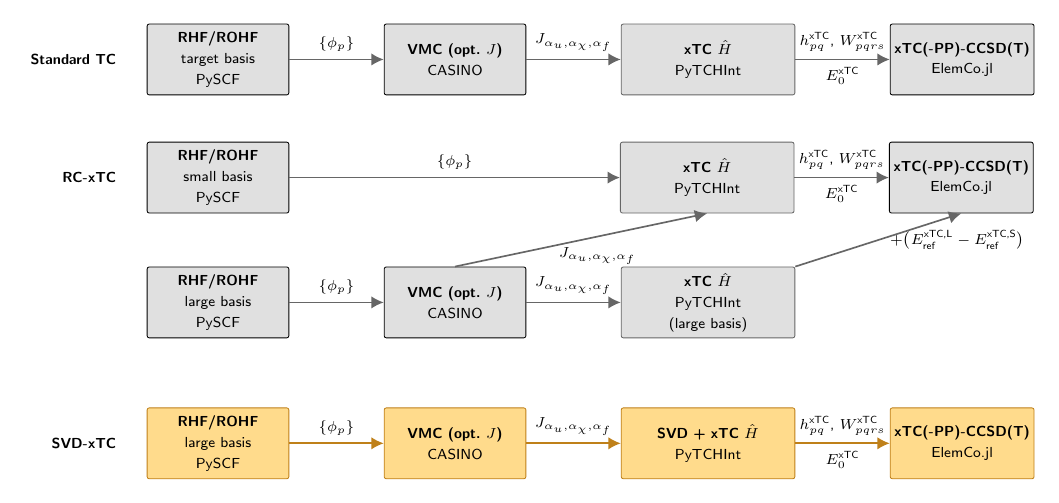}
    \caption{\label{fig:workflow}
    Three generations of TC workflows used in this work, one row per method variant. Boxes represent calculation stages and the codes employed; directed arrows indicate the flow of data between stages, with labels specifying the quantities transferred. \textbf{Standard xTC} (top): the RHF/ROHF reference is solved directly in the target basis. \textbf{RC-xTC} (middle): the correlation pipeline runs in the small basis (upper sub-row), while the lower sub-row uses the large-basis RHF/ROHF reference to optimise the Jastrow in VMC and to build a second, large-basis xTC Hamiltonian. The Jastrow is shared between the two PyTCHInt instances, and the difference of xTC reference energies, $E_{\text{ref}}^{\text{xTC,L}} - E_{\text{ref}}^{\text{xTC,S}}$, is added to the final xTC-CCSD(T) result. \textbf{SVD-xTC} (bottom): the RHF/ROHF reference is solved in the large basis and the virtual space is compressed inside PyTCHInt using the small-basis AO overlap matrices to define the target subspace; no additional small-basis HF is required.}
\end{figure*}

The workflow common to all calculations in this work is summarized in Fig.~\ref{fig:workflow}. Each stage is described in the following.

Reference orbitals are obtained from restricted Hartree--Fock (RHF) and restricted open-shell Hartree--Fock (ROHF) calculations for closed- and open-shell systems, respectively, using PySCF~\cite{sun2020}. We use augmented Dunning correlation-consistent basis sets, aug-cc-pV$x$Z, with $x=\mathrm{D,T,Q}$. For second-row elements, we use the corresponding aug-cc-pV($x$+d)Z variants. For two specific elements, Li and Na, for which core--valence correlation is non-negligible, we instead use the corresponding core--valence basis sets, aug-cc-pCV$x$Z. This convention is included implicitly in the AV$x$Z label throughout the Results section. We also performed all-electron calculations using the ANO family of basis sets, aug-ano-pV$x$Z, with the same cardinal numbers. Since the conclusions were similar to those obtained with the Dunning basis sets, we restrict the Results section to the Dunning basis sets for clarity. Details of the Dunning and ANO basis sets are provided in the Supplementary Material~\cite{interactivefigs}. 

For pseudopotential calculations, we use ccECPs together with the corresponding Dunning aug-cc-pV$x$Z basis sets~\cite{bennett2017,bennett2018}, following the transcorrelated treatment of pseudopotentials introduced in Ref.~\cite{simula2025ecp}. In this standard set, the first-row atoms and Na have a [He] core, while Si, P, S, and Cl have a [Ne] core; we refer to this set as Ne-core ccECPs. In a second set of calculations, referred to as He-core ccECPs, Si, P, S, and Cl are also described by [He]-core ccECPs, and aug-cc-pCV$x$Z basis sets are used for the second-row atoms. We also performed pseudopotential calculations using the non-augmented Dunning basis sets. These calculations led to reduced accuracy in the atomization energies, and the corresponding results are also reported in the Supplementary Material~\cite{interactivefigs}. 

For standard xTC calculations, the HF reference is solved directly in the target basis. For RC-xTC, HF references are computed in both a small and a large basis, and the additive reference-energy correction of Ref.~\cite{hauskrecht2026} is evaluated from their difference. For SVD-xTC, only a single HF calculation in the large basis is performed; the small basis enters solely through the AO overlap matrices, which determine the dimension and target subspace of the compressed virtual space. In all RC-xTC and SVD-xTC calculations reported below, the large basis used for the reference correction is AV5Z, except for Li, for which aug-cc-pCVQZ is used because no aug-cc-pCV5Z basis set is available.

For each system, the Jastrow parameters $\alpha_u, \alpha_\chi, \alpha_f$ are optimized in VMC using the CASINO package~\cite{needs2020}, with the single HF determinant trial wave function. The optimization minimizes the variance of the transcorrelated reference energy following the procedure of Ref.~\cite{haupt2023}. For standard xTC the trial wave function uses the target-basis HF orbitals; for RC-xTC and SVD-xTC it uses the large-basis (AV5Z) HF orbitals, so the same Jastrow is shared between RC-xTC and the corresponding SVD-xTC calculation (Fig.~\ref{fig:workflow}). We use cutoffs of $(L_u,L_\chi,L_f)=(4.5,1,2)$\,bohr and polynomial orders of $(k_u, k_\chi, k_f)=(8,8,3)$ in all calculations.

Given the optimized Jastrow factor and the reference orbitals, the xTC and xTC-PP correction integrals of Eqs.~\eqref{eq:T_pqrs}, \eqref{eq:F_pq}, and \eqref{eq:e0-xtc} are evaluated numerically with our in-house code PyTCHInt. The SVD virtual compression described above is implemented as a built-in preprocessing step in PyTCHInt.

The resulting non-TC and xTC second-quantized Hamiltonians are passed via the FCIDUMP format~\cite{Knowles89} or as Numpy binary files~\cite{Harris2020} to the ElemCo.jl package~\cite{elemcojl}, which performs the xTC-CCSD(T) (all-electron) and xTC-PP-CCSD(T) (with ECPs) calculations. In the results section we denote the full method as SVD-xTC(-PP)-CCSD(T); the (-PP) is present when ccECPs are used and absent in the all-electron case. For open-shell systems, ElemCo.jl solves the spin-unrestricted CCSD equations based on the ROHF determinant, and the perturbative triples are evaluated in the $\Lambda$CCSD(T) formulation~\cite{taube2008}. We thus use $\Lambda$UCCSD(T) for open-shell and $\Lambda$CCSD(T) for closed-shell systems. For comparison, we also perform non-TC CCSD(T) calculations with Molpro in the same basis sets as the corresponding xTC calculations, including the core--valence basis sets for Li and Na. These calculations use ROHF orbitals with the UCCSD(T) method for open-shell systems and CCSD(T) for closed-shell systems. They therefore employ the standard (T) correction. To assess the effect of the two triples definitions, we repeated the non-TC calculations with ElemCo.jl in AVDZ, AVTZ, and AVQZ, using both the standard (T) and the $\Lambda$CCSD(T) correction on the same non-TC Hamiltonians, and find that the choice of the triples definition changes the atomization energies only slightly and thus does not affect the discussion below.

For a molecule $M$ composed of atoms $A$ with stoichiometric coefficients $n_A$, the atomization energy (AE) is
\begin{equation}
  \label{eq:atomization}
  \mathrm{AE}_M = \sum_{A} n_A\, E_A - E_M,
\end{equation}
where $E_M$ and $E_A$ are the absolute molecular and atomic total energies. Defining the per-molecule signed error against a reference,
$\Delta_M = \mathrm{AE}_M^{\text{method}} - \mathrm{AE}_M^{\text{ref}}$, we report three error metrics over the set of $N$ molecules:
\begin{equation}
  \label{eq:err-metrics}
  \mathrm{MAE} = \tfrac{1}{N}\sum_M |\Delta_M|,
  \quad
  \mathrm{RMSE} = \sqrt{\tfrac{1}{N}\sum_M \Delta_M^2},
  \quad
  \mathrm{MaxE} = \max_M |\Delta_M|.
\end{equation}

The G2-1 benchmark set introduced above is more precisely the G2/55 subset of the Gaussian-2 test set of Curtiss et al.~\cite{curtiss1991}. Throughout the paper, all method-vs-method comparisons are made at the level of the equilibrium atomization energy $D_e$ against the SHCI+PBE+CV values of Ref.~\cite{Yao2020}, extrapolated to both FCI limit and complete basis set (CBS) limit. SHCI+PBE+CV combines an extrapolated frozen-core SHCI total $D_e$ at the CBS limit with a PBE-based correction for residual basis-set incompleteness and a core-valence (CV) correction. The CV correction is added only to the SHCI value, because the underlying SHCI is frozen-core; the all-electron TC explicitly correlates the core electrons, while the ECP TC inherits core-valence correlation implicitly through the ccECP construction. Throughout the remainder of the paper we refer to this reference uniformly as SHCI+PBE+CV, with the CBS extrapolation implicit. 

When comparing against the experimental atomization energy from the same compilation, theoretical $D_e$ values are converted to $D_0$ by adding zero-point energy (ZPE) and scalar-relativistic / spin-orbit (SR\_SO) correction. The per-molecule values of CV, the ZPE, and SR\_SO correction used throughout this work are taken from the compilation of Ref.~\cite{Yao2020}, which collects them from the composite-energy studies of Feller, Peterson, and Dixon~\cite{10.1063/1.478747,10.1063/1.3008061}. Per-basis, per-molecule discrepancies and the full basis-set convergence of error metrics, as well as timings of all the calculations done here, are hosted as interactive plotly figures~\cite{interactivefigs}.

\section{Results}
\begin{figure*}[htbp]
  \centering
  \includegraphics[width=\textwidth]{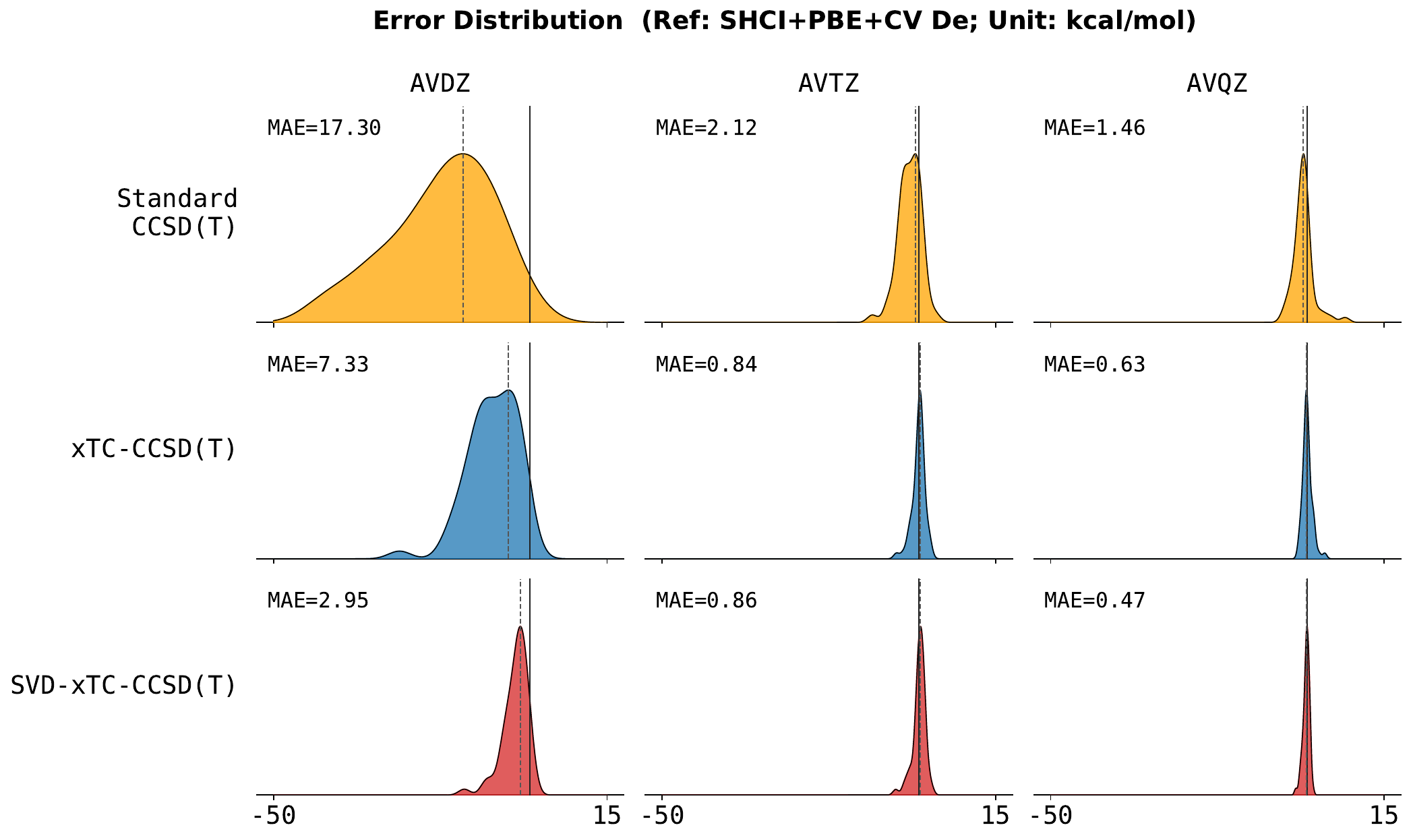}
  \caption{\label{fig:kde_standard_vs_tc_PBE}Gaussian kernel-density estimate of atomization-energy errors relative to SHCI+PBE+CV $D_e$ (kcal$\cdot$mol$^{-1}$) for standard CCSD(T) (orange), xTC-CCSD(T) (blue) and SVD-xTC-CCSD(T) (red) at AVxZ with $x =$ D, T, Q. Each KDE is constructed using Gaussian kernels, with the bandwidth determined by Scott's rule. The solid vertical line marks zero error; the dashed line marks the KDE mode (distribution peak). Annotated MAE values denote the corresponding mean absolute errors.}
\end{figure*}

To assess the performance of the transcorrelated approach on the G2-1 dataset, we first compare xTC-CCSD(T) and SVD-xTC-CCSD(T) with conventional CCSD(T) in Fig.~\ref{fig:kde_standard_vs_tc_PBE}. We summarize the error distributions using kernel-density estimates (KDEs), constructed by placing a Gaussian function at each data point and summing these functions to obtain a smooth estimate of the underlying distribution, and report the MAEs.

Across all three basis sets, xTC-CCSD(T) systematically narrows the error distribution and reduces the mean absolute error (MAE). The improvement is particularly pronounced in AVDZ, where basis-set incompleteness errors are largest: the MAE is reduced from 17.30 to 7.33 kcal$\cdot$mol$^{-1}$. In AVTZ and AVQZ, xTC-CCSD(T) reaches chemical-accuracy MAEs of 0.84 and 0.63 kcal$\cdot$mol$^{-1}$, respectively, whereas conventional CCSD(T) gives larger MAEs of 2.12 and 1.46 kcal$\cdot$mol$^{-1}$. The progressive narrowing of the xTC-CCSD(T) error distribution from AVDZ to AVQZ also indicates a clear and systematic convergence with increasing basis-set size.

SVD-xTC-CCSD(T) in Fig.~\ref{fig:kde_standard_vs_tc_PBE} reduces the MAEs even more, especially in AVDZ, where it reaches $2.95$ kcal$\cdot$mol$^{-1}$. In AVTZ, SVD-xTC-CCSD(T) is chemically accurate with MAE of 0.86 kcal$\cdot$mol$^{-1}$. In AVQZ, SVD-xTC achieves the lowest MAE of $0.47$ kcal$\cdot$mol$^{-1}$, outperforming standard xTC-CCSD(T), and yields the narrowest distribution of the three methods.

\begin{figure*}[htbp]
  \centering
  \includegraphics[width=\textwidth]{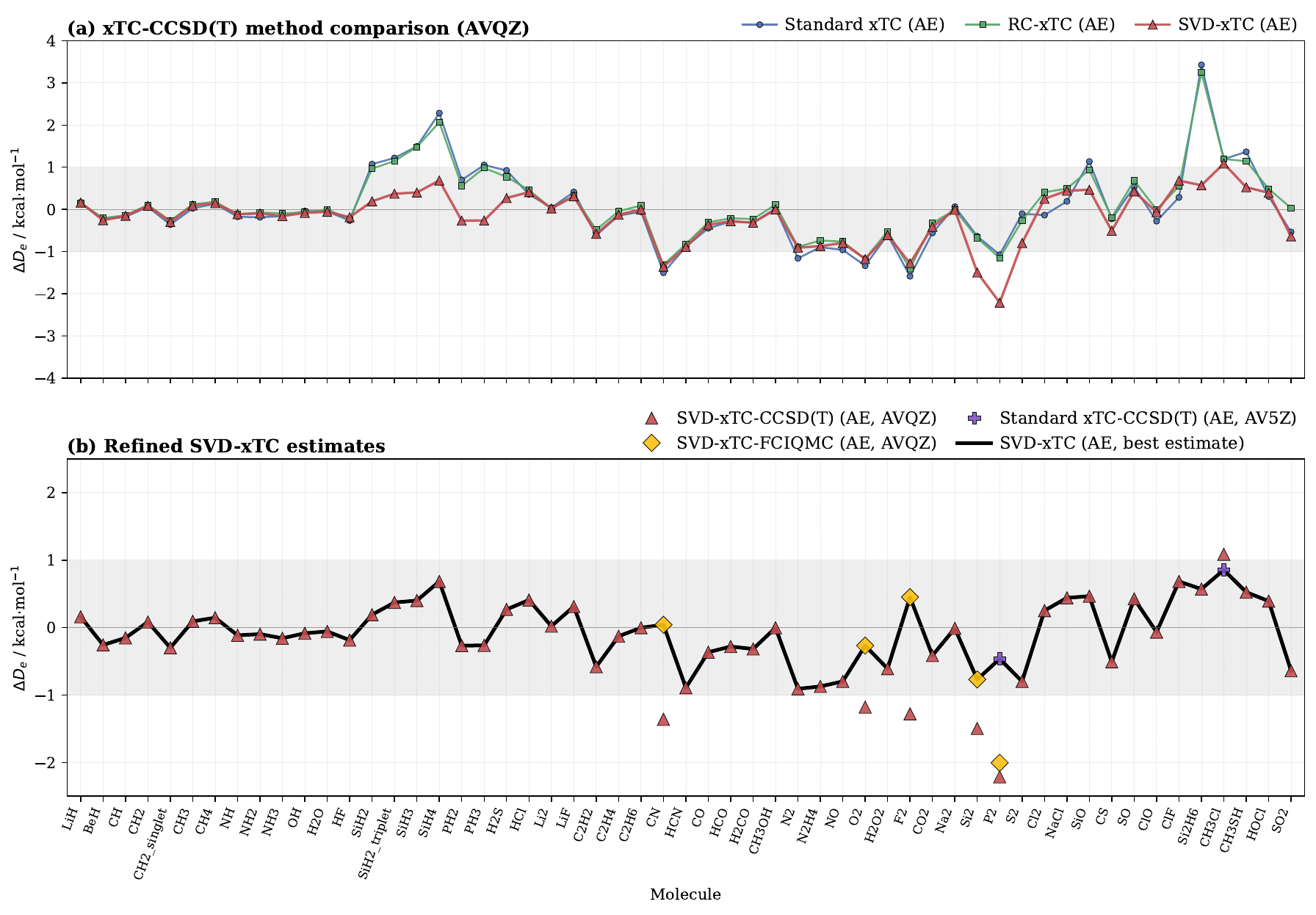}
  \caption{\label{fig:molecule_discrepancy}
    Per-molecule atomization-energy discrepancy $\Delta D_e$ vs SHCI+PBE+CV. 
    (a) Comparison of standard xTC-, RC-xTC-, and SVD-xTC-CCSD(T) at the AVQZ basis set.
    (b) SVD-xTC results on a zoomed scale: SVD-xTC-CCSD(T) at AVQZ, SVD-xTC-FCIQMC at AVQZ for CN, O$_2$, F$_2$, Si$_2$ and P$_2$, and xTC-CCSD(T) at AV5Z for P$_2$ and CH$_3$Cl.
    The black line traces the per-molecule best estimate among all-electron xTC results in this work. Grey band marks chemical accuracy. AE denotes the all-electron calculation.}
\end{figure*}

In the upper panel (a) of Fig.~\ref{fig:molecule_discrepancy}, we show the per-molecule AE discrepancies of standard, RC- and SVD-xTC-CCSD(T) relative to SHCI+PBE+CV in the largest basis set considered here, AVQZ. RC-xTC is comparable to standard xTC-CCSD(T), because the xTC reference energy is already nearly converged in AVQZ and the reference correction is therefore marginal. These methods have positive outliers with Si-containing molecules (triplet SiH$_2$, SiH$_3$, SiH$_4$, and Si$_2$H$_6$). The tendency of standard and RC-xTC to give higher AEs and thus also positive outliers is unclear; possible reasons include overcorrelation of the molecules or unbalanced xTC description between the atoms and molecules. The latter is feasible given the chemical differences between atomic and covalently bonded orbitals. However, SVD-xTC is found to fix this issue. This is possibly because of the improved occupied-virtual interactions, as discussed in the Theory section. For explicit analysis, we split the AE of SiH$_4$ (Eq.~\ref{eq:atomization}) into a reference contribution and a correlation contribution, $\mathrm{AE}_{\text{corr}} = \sum_A n_A E_{\text{corr},A} - E_{\text{corr},M}$, where $E_{\text{corr}}$ is the xTC-CCSD(T) correlation energy. The reference contribution is identical in RC-xTC and SVD-xTC. The correlation energies, however, differ: the SVD-xTC correlation energies are 8.13 kcal$\cdot$mol$^{-1}$ and 6.74 kcal$\cdot$mol$^{-1}$ lower than those of RC-xTC for the Si atom and the SiH$_4$ molecule, respectively. As a result, the correlation contribution to the AE of SiH$_4$ decreases by 1.39 kcal$\cdot$mol$^{-1}$, which reduces its deviation from SHCI+PBE+CV from 1.87 to 0.48 kcal$\cdot$mol$^{-1}$. Although SVD-xTC has negative outliers not seen with RC-xTC or standard xTC (discussed below), we claim that the tendency to overshoot AEs is the reason for not seeing these outliers with standard and RC-xTC. 

The outliers of SVD-xTC-CCSD(T) method are with molecules $\mathrm{CN}$, $\mathrm{O}_2$, $\mathrm{F}_2$, $\mathrm{Si}_2$, $\mathrm{P}_2$, and $\mathrm{CH}_3\mathrm{Cl}$  in AVQZ. To test if a stronger multireference character of these molecules is behind the error, we performed SVD-xTC-FCIQMC \cite{luo2018combining} calculations to determine whether these outliers arise primarily from basis-set incompleteness or from the approximate correlation treatment in CCSD(T). The corresponding results are shown in the lower panel of Fig.~\ref{fig:molecule_discrepancy}.

For the atomic SVD-xTC-FCIQMC calculations, we used a walker population of $10^8$ that was sufficient to converge the energies to within $0.1$ millihartree. For the molecular calculations, we used energy estimates with walker populations of $10^7$, $10^8$, and $5\times10^8$ to extrapolate to the infinite-walker limit\cite{haupt2023,hosseini2024combining}. The results clearly indicate that the errors for $\mathrm{CN}$, $\mathrm{O}_2$, $\mathrm{F}_2$, and $\mathrm{Si}_2$ arise predominantly from the remaining correlation error in CCSD(T), whereas the FCIQMC treatment reduces the error for $\mathrm{P}_2$ only marginally. The xTC-CCSD(T) result for $\mathrm{P}_2$ obtained with the AV5Z basis set suggests that its error originates mainly from basis-set incompleteness. For $\mathrm{CH}_3\mathrm{Cl}$, an SVD-xTC-FCIQMC result is not available because of its prohibitively large orbital space; nevertheless, increasing the basis set to AV5Z brings the xTC-CCSD(T) result within chemical accuracy.

\begin{figure*}[htbp]
  \centering
  \includegraphics[width=\textwidth]{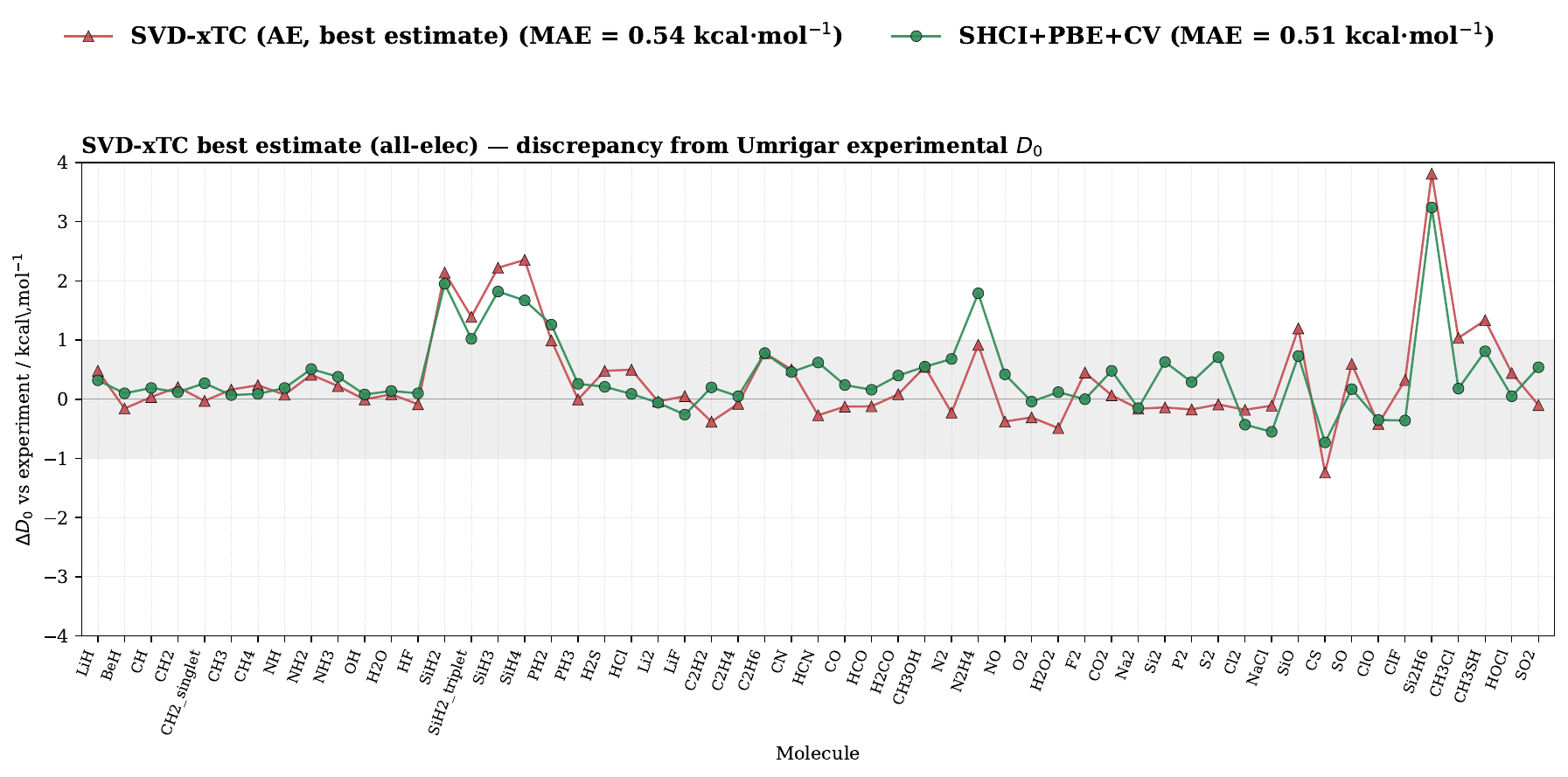}
  \caption{\label{fig:avqz_vs_experiment}
    Per-molecule atomization-energy discrepancy $\Delta D_0$ vs the Umrigar experimental $D_0$~\cite{Yao2020}, for the SVD-xTC (all-electron) best estimate (cf.\ Fig.~\ref{fig:molecule_discrepancy}) and SHCI+PBE+CV. Theoretical $D_0$ values include ZPE and scalar-relativistic / spin-orbit (SR\_SO) corrections on top of $D_e$; SHCI additionally includes the core-valence (CV) correction (frozen-core), whereas the all-electron TC already contains core correlation. Grey band marks chemical accuracy.}
\end{figure*}

Although SHCI+PBE+CV should, in principle, provide highly accurate reference values, it is not guaranteed to be exact. The accuracy of SHCI depends on two extrapolations~\cite{sharma2017}: extrapolation to the FCI limit and extrapolation to the CBS limit. Therefore, in Fig.~\ref{fig:avqz_vs_experiment}, we also compare the SVD-xTC best estimate (the AVQZ values with the per-molecule FCIQMC and AV5Z substitutions of Fig.~\ref{fig:molecule_discrepancy}) and SHCI+PBE+CV against the experimental atomization energies. This produces nearly identical AE patterns between SVD-xTC and SHCI+PBE+CV, with very similar MAEs: 0.54 (SVD-xTC) and 0.51 (SHCI+PBE+CV) kcal$\cdot$mol$^{-1}$. The agreement between the two methods hints at high accuracy; the methods differ by the approach to capture correlation and by converging basis set, but both target the FCI values at CBS. 


In the comparison between theory and experiment, there are intrinsic uncertainties in the experimental reference values, together with possible errors in the corrections applied to the theoretical values, such as ZPE and SR\_SO. The authors of the SHCI study on the G2-1 set~\cite{Yao2020} noted that this makes the comparison difficult and highlights the need for high-quality quantum-chemical benchmark values from other theoretical and computational methods. 
Our results support the view that SVD-xTC and SHCI+PBE+CV provide a better reference to theoretical methods than experiment for the G2-1 set. However, since the errors of SHCI+PBE+CV, experiment, and SVD-xTC-CCSD(T) are all within 1 kcal$\cdot$mol$^{-1}$, it remains difficult to draw a definitive conclusion on the correct values.

The results above suggest that SVD-xTC methods are an accurate and practical way for quantum chemical simulations. However, especially for the systems with second-row atoms, the Jastrow factor optimization gets computationally expensive due to large variances of the VMC wave functions for large $Z$ nuclei. These problems can be circumvented with the use of pseudopotentials. Pseudopotentials include approximately the core-valence and scalar relativistic corrections, but introduce an error of unknown magnitude. The G2/55 dataset offers a powerful way to inspect the errors of the pseudopotential approximation in the SVD-xTC method. 

\begin{figure*}[t]
  \centering
  \includegraphics[width=\textwidth]{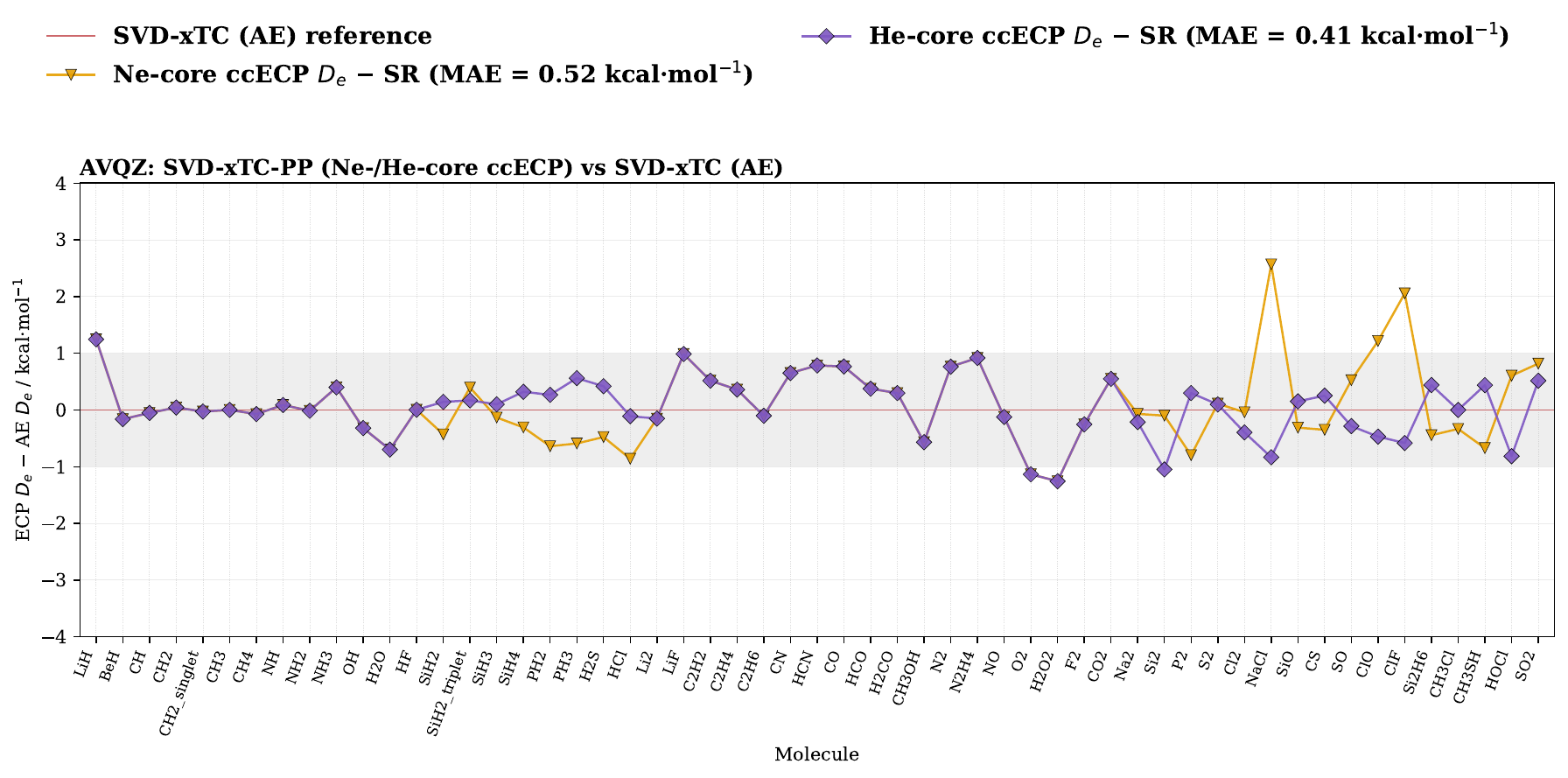}
  \caption{\label{fig:avqz_ae_vs_ecp}
    Per-molecule $D_e$ difference of SVD-xTC-PP-CCSD(T) from the all-electron SVD-xTC-CCSD(T) reference (zero line) at AVQZ. Two ECP curves are shown, both with the per-molecule scalar-relativistic (SR) contribution subtracted (the convention used when comparing to non-relativistic SHCI+PBE+CV): with Ne-core ccECPs, and with He-core ccECPs. Grey band marks chemical accuracy.}
\end{figure*}

Figure \ref{fig:avqz_ae_vs_ecp} compares AE results between the all-electron SVD-xTC-CCSD(T) and SVD-xTC-PP-CCSD(T). The comparison is done in AVQZ (for the He-core ccECPs, aug-cc-pCVQZ on the second-row atoms). We show the ECP $D_e$ with the per-molecule SR contribution subtracted (the convention used when comparing to non-relativistic SHCI+PBE+CV), both with Ne-core and with He-core ccECPs.

We can see the inherent error in the ECP approximation, which is on average 0.52 $\mathrm{kcal\cdot mol^{-1}}$ for the Ne-core ccECPs and 0.41 $\mathrm{kcal\cdot mol^{-1}}$ for the He-core ccECPs. These results make the SVD-xTC-PP methods highly reliable for accurate simulations, in particular with He-core ccECPs for the second-row atoms.

When Ne-core ccECPs are used, the outliers among systems containing second-row atoms 
are NaCl, ClF, and ClO (Fig.~\ref{fig:avqz_ae_vs_ecp}). The He-core ccECPs fixes all three. It remains unclear whether Ne-core pseudopotential has an error in the pseudopotential itself, or if the larger ccECP radius of the Ne-core ccECPs leads to larger errors due to TC approximations done with the pseudopotential commutators. Commutators beyond second order are tested with no effect, and 3-body pseudopotential commutators are outside the scope of this work.

\begin{figure*}[htbp]
  \centering
  \includegraphics[width=\textwidth]{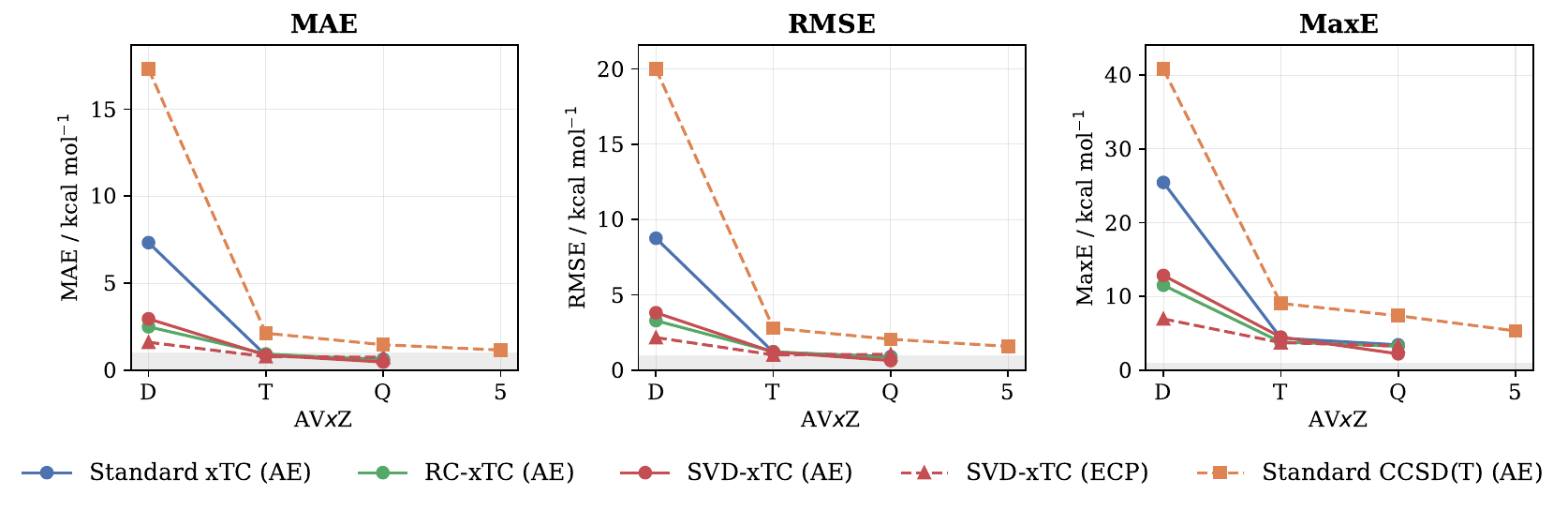}
  \caption{\label{fig:ae_method_comparison}
    Error metrics (MAE, RMSE, MaxE in kcal$\cdot$mol$^{-1}$) of CCSD(T) atomization energies vs SHCI+PBE+CV, for Standard CCSD(T), xTC-, RC-xTC-, and SVD-xTC-CCSD(T) (all-electron), and for SVD-xTC-PP-CCSD(T) with Ne-core ccECPs on AV$x$Z. Grey band marks chemical accuracy ($\le 1$\,kcal$\cdot$mol$^{-1}$).}
\end{figure*}

In Fig.~\ref{fig:ae_method_comparison}, we compare the error metrics MAE, RMSE, and MaxE against SHCI+PBE+CV for standard CCSD(T), xTC-, RC-xTC-, and SVD-xTC(-PP)-CCSD(T) (all-electron / Ne-core ccECP) on AV$x$Z, with $x=$ D, T, Q. The corresponding values are given in Table~\ref{tab:vs_shci}. For clarity, we do not plot the He-core ccECP metrics, as they are very close to the all-electron ones (see Table~\ref{tab:vs_shci}).
Overall the SVD-xTC method provides similar accuracy to RC-xTC. In AVDZ, it offers comparable though slightly worse accuracy, while in larger basis sets it is better, with absolute differences in error metrics clearly under $1$ kcal$\cdot$mol$^{-1}$ in all basis sets. We also note that SVD-xTC gives lower total energies, likely because the improved occupied-virtual correlation interactions in the second-quantized xTC Hamiltonian.  Given that SVD-xTC is more practical and is no longer a composite method, we recommend it as the preferred approach for TC applications.  

The error metrics with Ne-core ccECPs are very interesting. ccECP errors are by far the lowest in AVDZ, and the error metrics are nearly identical between AVTZ and AVQZ (see also Table~\ref{tab:vs_shci}). Hence, the basis set convergence is much faster with Ne-core ccECPs and converges in AVTZ. The reason likely lies in the more targeted Jastrow optimization: in the all-electron case, the Jastrow must describe correlation in the core region, which is mostly irrelevant to chemical bonding, whilst at the same time contributing a considerable amount to the total correlation energy. 
In the ECP case the Jastrow only has to describe valence correlation, so the optimization targets the correlation relevant to bonding, yielding a more effective Jastrow. Also, there are no electron-nuclear cusps to be described with ECPs, potentially further accelerating single-particle basis-set convergence. The considerations are further supported by the He-core ccECP error metrics in Table~\ref{tab:vs_shci}: as only 1s orbitals of 2nd-row atoms are included in the ECP, the simulation is close to all electron and the error metrics are very similar to the all-electron case. As shown in Table \ref{tab:vs_shci}, we see with Ne-core ccECPs that the MAE and RMSE in the xTC-CCSD(T) atomization energies are converged in AVTZ, and do not improve in going to AVQZ, unlike the all-electron case where a further reduction in the error statistics is obtained. This means that we have converged the Ne-core ccECP calculation of the atomization energies close to its intrinsic accuracy of 0.6-0.8 kcal/mol already at the AVTZ level.

\begin{table}[h]
  \centering
  \small
  \caption{\label{tab:vs_shci}
    Error metrics (kcal\,mol$^{-1}$) of computed $D_e$ atomization energies vs the SHCI+PBE+CV reference (= $D_e^{\text{SHCI+PBE}} + \mathrm{CV}$) on the G2/55 set, and (bottom section) vs the Umrigar experimental $D_0$. The best-estimate row combines AVQZ values with molecule-specific FCIQMC and AV5Z substitutions (see Fig.~\ref{fig:molecule_discrepancy}); the substituted molecules are listed in note $a$. All error metrics are evaluated over all 55 molecules of the set.}
  \begin{threeparttable}
  \begin{tabular}{l l r r r}
    \toprule
    Method & Basis & MAE & RMSE & MaxE \\
    \midrule
    Standard CCSD(T) (AE) & AVDZ & 17.30 & 19.98 & 40.81 \\
     & AVTZ & 2.12 & 2.79 & 9.04 \\
     & AVQZ & 1.46 & 2.05 & 7.38 \\
    \midrule
    Standard xTC (AE) & AVDZ & 7.33 & 8.75 & 25.45 \\
     & AVTZ & 0.84 & 1.19 & 4.35 \\
     & AVQZ & 0.63 & 0.89 & 3.43 \\
    \midrule
    RC-xTC (AE) & AVDZ & 2.49 & 3.29 & 11.51 \\
     & AVTZ & 0.93 & 1.22 & 3.80 \\
     & AVQZ & 0.57 & 0.82 & 3.25 \\
    \midrule
    SVD-xTC (AE) & AVDZ & 2.95 & 3.81 & 12.84 \\
     & AVTZ & 0.86 & 1.20 & 4.48 \\
     & AVQZ & 0.47 & 0.64 & 2.21 \\
    \midrule
    SVD-xTC + Ne-core ccECP & AVDZ & 1.58 & 2.11 & 6.65 \\
     & AVTZ & 0.63 & 0.85 & 3.48 \\
     & AVQZ & 0.65 & 0.99 & 3.01 \\
    \midrule
    SVD-xTC + He-core ccECP & AVDZ & 2.43 & 3.42 & 12.41 \\
     & AVTZ & 0.73 & 1.19 & 5.00 \\
     & AVQZ & 0.54 & 0.80 & 2.54 \\
    \midrule
    \multicolumn{5}{l}{\textit{vs Umrigar experimental $D_0$}} \\
    SVD-xTC-CCSD(T) (AE) & AVQZ & 0.62 & 0.97 & 3.81 \\
    SVD-xTC (AE, best estimate)\tnote{a} & AVQZ/AV5Z & 0.54 & 0.89 & 3.81 \\
    SVD-xTC-PP-CCSD(T) (Ne-core ccECP) & AVQZ & 0.86 & 1.18 & 3.37 \\
    SHCI+PBE+CV & - & 0.51 & 0.78 & 3.24 \\
    \bottomrule
  \end{tabular}
  \begin{tablenotes}
    \footnotesize
    \item[a] SVD-xTC-CCSD(T)/AVQZ for 49 of the 55 molecules. Six molecules are substituted: CN, O$_2$, F$_2$, and Si$_2$ by SVD-xTC-FCIQMC/AVQZ, and P$_2$ and CH$_3$Cl by xTC-CCSD(T)/AV5Z.
  \end{tablenotes}
  \end{threeparttable}
\end{table}

\begin{figure}[h]
  \centering
  \includegraphics[width=0.8\textwidth]{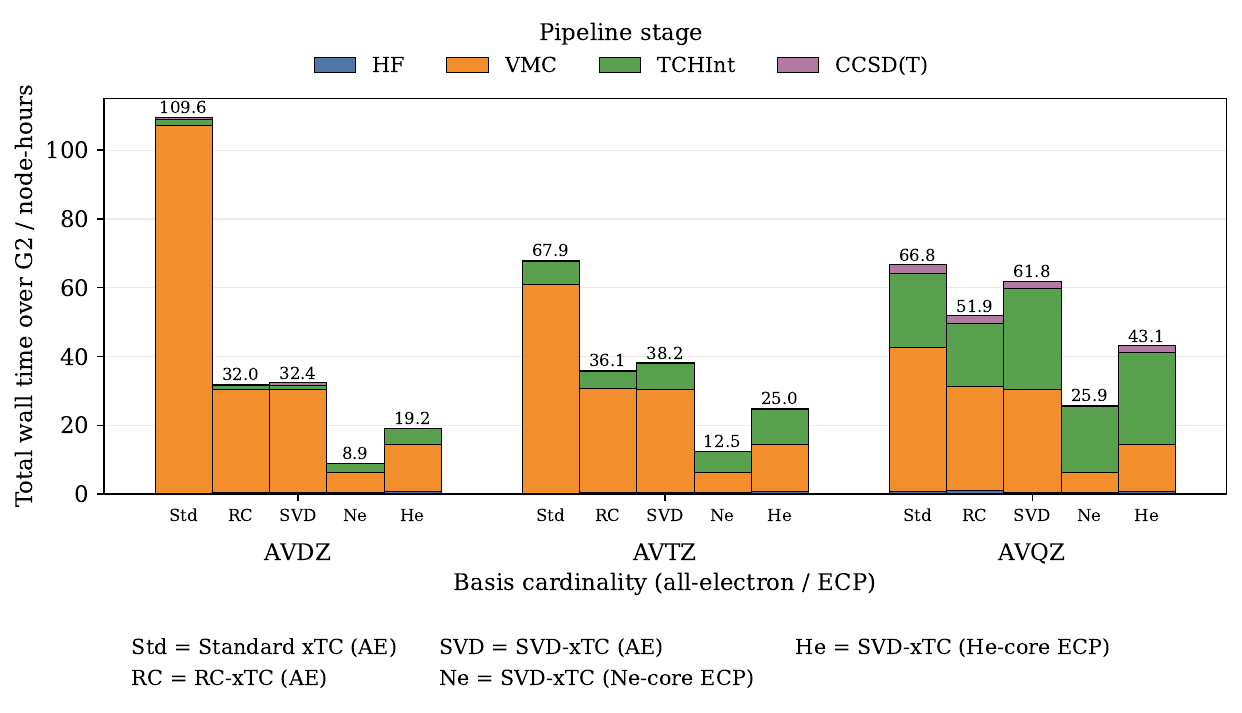}
  \caption{\label{fig:timing_comparison}
    Aggregate wall time over the G2-1 set (node-hours) for Standard xTC, RC-xTC, SVD-xTC-CCSD(T) (all-electron), and SVD-xTC-PP-CCSD(T) (Ne-core and He-core ccECPs), with AV$x$Z at $x=$ D, T, Q. The timings were obtained on AMD 128-core compute nodes, primarily equipped with AMD EPYC 7H12 64-Core processors, with a small number of calculations performed on nodes equipped with AMD EPYC 7763 and 9554 64-Core processors.}
\end{figure}

Figure~\ref{fig:timing_comparison} compares the total computational cost of the different xTC methods and basis sets over the G2-1 set, expressed in node-hours. The calculations were performed on AMD 128-core nodes (mostly equipped with two AMD EPYC 7H12 64-core processors; see the caption of Fig.~\ref{fig:timing_comparison}). Although the nodes were not identical and their performance may have fluctuated, these results still provide a rough estimate of the computational cost of the TC workflow. SVD-xTC-CCSD(T) requires 38.2 node-hours at AVTZ and 61.8 node-hours at AVQZ. As the basis set size increases, the dominant cost starts to shift from VMC optimization to the numerical evaluation of the xTC corrections and the CCSD(T) calculation. VMC optimization is performed in real space, and the basis set enters only through the HF orbitals in the Slater-Jastrow trial wave function. Larger basis sets yield more accurate HF orbitals and thus reduce the variance of the local energy, so that fewer VMC steps are needed to achieve the same statistical accuracy~\cite{haupt2023}. Consequently, the VMC optimization cost is highest for the smallest basis set in standard xTC calculations. With Ne-core ccECPs, the VMC optimization is relatively inexpensive, and xTC integration becomes the computational bottleneck for larger basis sets because the spherical projections required by the ECPs are more expensive to evaluate~\cite{simula2025ecp}. With He-core ccECPs, the cost of the VMC optimization falls between those of the all-electron and Ne-core ccECP calculations. The total computational cost of evaluating the atomization energies of the G2 set using SVD-xTC-PP-CCSD(T) with Ne-core ccECPs at AVDZ is only 8.9 node-hours, which is particularly low considering the small MAE of approximately 1.5\,kcal$\cdot$mol$^{-1}$ relative to SHCI+PBE+CV reference values.

\section{Conclusions and Outlook}

In this work, we developed the reference correction based on SVD (SVD-xTC) scheme for transcorrelated theory. SVD-xTC(-PP) retains the highly accurate reference state throughout the TC workflow, and compresses the virtual space from the large-basis SCF calculation by projecting onto a small-basis AO space. The method, using CCSD(T) as the correlation solver on the SVD-xTC(-PP) Hamiltonians, was benchmarked against the near-exact SHCI+PBE+CV and experimental atomization energies for the G2-1 set.

All three TC variants we tested -- standard, RC-xTC, and SVD-xTC(-PP)-CCSD(T) -- reach chemical accuracy in atomization energies against both SHCI+PBE+CV (at AVTZ) and experiment (at AVQZ). This is a direct consequence of the wave-function compactification: by pulling correlation into the occupied space, the TC similarity transformation leaves so little for the post-Hartree--Fock solver that CCSD(T) alone -- previously considered insufficient for quantitative accuracy on the G2 set~\cite{Yao2020} -- suffices for chemically accurate atomization energies of first- and second-row molecules. The timings reported in this work can be further reduced through grid-pruning and density-fitting schemes, both under active development, and are therefore expected to compare favorably with those of alternative quantum-chemical approaches.

We found SVD-xTC(-PP)-CCSD(T) to have similar acceleration in basis-set convergence of atomization energies as the RC-xTC scheme reported earlier~\cite{hauskrecht2026}. Moreover, for molecules containing second-row elements, the errors of SVD-xTC-CCSD(T) relative to SHCI+PBE+CV decreased to within the chemical-accuracy threshold, whereas the standard and RC-xTC schemes overshot the reference values in some cases with the largest basis set. This convergence behavior, together with SVD-xTC's simpler workflow (a single SCF calculation rather than two), and its improved occupied-virtual interactions from the large-basis reference orbitals, make SVD-xTC our recommended TC method going forward.

We measured the ECP errors in SVD-xTC-PP-CCSD(T) both with He- and Ne-core ccECPs, which were on average $0.41$\,kcal$\cdot$mol$^{-1}$ and $0.52$\,kcal$\cdot$mol$^{-1}$. Use of He-core ccECPs fixed all the outliers with Ne-core ccECPs for molecules with 2nd-row atoms. The Jastrow optimization with ECPs is likely more effective because it directly targets the bonding orbitals: with Ne-core ccECPs, the results converged at AVTZ, while with He-core ccECPs the results and convergence were close to all-electron ones. Hence He-core ccECPs represent for 2nd-row atoms a compromise between accuracy and cost: VMC optimization still is cheaper than all-electron, with similar accuracy. 

Two directions remain open for future work. First, pushing TC to chemical accuracy already in the AVDZ basis, where the ECP results give a hint on how to proceed: the AVDZ SVD-xTC-PP-CCSD(T) with Ne-core ccECPs attains a mean absolute error against CBS SHCI+PBE+CV of only $1.58$\,kcal$\cdot$mol$^{-1}$, at only $8.9$ node-hours over the full G2 set. Second, combining the favorable Jastrow optimization seen with ECPs with all-electron TC workflows; recently developed deterministic optimization of Jastrow factors~\cite{filip2025jastrow}, possibly with a frozen-core approximation, could be a way forward.

\begin{acknowledgement}
The authors gratefully acknowledge generous funding from the Max Planck Society.
\end{acknowledgement}


\section*{Data and Software Availability}
The data and parameter settings underlying this study are available in the published article and at https://simulak1.github.io/tc-g2-data/. All software used to generate the results is publicly available, except for PyTCHInt, which will be made publicly available in an upcoming release.


\clearpage
\bibliography{refs}

@article{Yao2020,
  author  = {Yao, Yuan and Giner, Emmanuel and Li, Junhao and Toulouse, Julien and Umrigar, C. J.},
  title   = {Almost exact energies for the {Gaussian-2} set with the semistochastic heat-bath configuration interaction method},
  journal = {J. Chem. Phys.},
  volume  = {153},
  number  = {12},
  pages   = {124117},
  year    = {2020},
  doi     = {10.1063/5.0018577},
  note    = {\href{https://doi.org/10.1063/5.0018577}{doi:10.1063/5.0018577}}
}

@misc{interactivefigs,
  author       = {Simula, Kristoffer and Cheng, Yifan},
  title        = {Interactive plotly figures: xTC atomization energies vs SHCI+PBE+CV},
  year         = {2026},
  howpublished = {Online collection, GitHub Pages},
  note         = {\url{https://simulak1.github.io/tc-g2-data/}}
}

@article{annaberdiyev2021,
  title     = {Cohesion and excitations of diamond-structure silicon by quantum Monte Carlo: Benchmarks and control of systematic biases},
  author    = {Annaberdiyev, Abdulgani and Wang, Guangming and Melton, Cody A. and Bennett, M. Chandler and Mitas, Lubos},
  journal   = {Phys. Rev. B},
  volume    = {103},
  issue     = {20},
  pages     = {205206},
  numpages  = {13},
  year      = {2021},
  month     = {May},
  publisher = {American Physical Society},
  doi       = {10.1103/PhysRevB.103.205206},
  url       = {https://link.aps.org/doi/10.1103/PhysRevB.103.205206}
}

@article{kats2024,
  author   = {Kats, Daniel and Christlmaier, Evelin M. C. and Schraivogel, Thomas and Alavi, Ali},
  title    = {Orbital optimisation in xTC transcorrelated methods},
  journal  = {Faraday Discuss.},
  volume   = {254},
  pages    = {382--401},
  year     = {2024},
  doi      = {10.1039/D4FD00036F},
  url      = {https://doi.org/10.1039/D4FD00036F}
}

@article{liao2021,
  title     = {Towards efficient and accurate ab initio solutions to periodic systems via transcorrelation and coupled cluster theory},
  author    = {Liao, Ke and Schraivogel, Thomas and Luo, Hongjun and Kats, Daniel and Alavi, Ali},
  journal   = {Phys. Rev. Res.},
  volume    = {3},
  issue     = {3},
  pages     = {033072},
  numpages  = {10},
  year      = {2021},
  month     = {Jul},
  publisher = {American Physical Society},
  doi       = {10.1103/PhysRevResearch.3.033072},
  url       = {https://link.aps.org/doi/10.1103/PhysRevResearch.3.033072}
}

@article{schraivogel2021,
  author  = {Schraivogel, Thomas and Cohen, Aron J. and Alavi, Ali and Kats, Daniel},
  title   = {Transcorrelated coupled cluster methods},
  journal = {The Journal of Chemical Physics},
  volume  = {155},
  number  = {19},
  pages   = {191101},
  year    = {2021},
  month   = {11},
  issn    = {0021-9606},
  doi     = {10.1063/5.0072495},
  url     = {https://doi.org/10.1063/5.0072495}
}

@article{liao2023dmrg,
  author  = {Liao, Ke and Zhai, Huanchen and Christlmaier, Evelin Martine Corvid and Schraivogel, Thomas and Ríos, Pablo López and Kats, Daniel and Alavi, Ali},
  doi     = {10.1021/acs.jctc.2c01207},
  journal = {Journal of Chemical Theory and Computation},
  note    = {PMID: 36912635},
  number  = {6},
  pages   = {1734–1743},
  title   = {Density Matrix Renormalization Group for Transcorrelated Hamiltonians: Ground and Excited States in Molecules},
  url     = { 
             
             https://doi.org/10.1021/acs.jctc.2c01207
             
             
             
             },
  volume  = {19},
  year    = {2023}
}

@article{haupt2025,
  author  = {Haupt, J. Philip and Christlmaier, Evelin M. C. and Lopez Ríos, Pablo and Bogdanov, Nikolay A. and Kats, Daniel and Alavi, Ali},
  title   = {Transcorrelated methods for multireference problems},
  journal = {The Journal of Chemical Physics},
  volume  = {163},
  number  = {14},
  pages   = {144113},
  year    = {2025},
  month   = {10},
  issn    = {0021-9606},
  doi     = {10.1063/5.0282673},
  url     = {https://doi.org/10.1063/5.0282673}
}

@article{drummond2004,
  title     = {Jastrow correlation factor for atoms, molecules, and solids},
  author    = {Drummond, N. D. and Towler, M. D. and Needs, R. J.},
  journal   = {Phys. Rev. B},
  volume    = {70},
  issue     = {23},
  pages     = {235119},
  numpages  = {11},
  year      = {2004},
  month     = {Dec},
  publisher = {American Physical Society},
  doi       = {10.1103/PhysRevB.70.235119},
  url       = {https://link.aps.org/doi/10.1103/PhysRevB.70.235119}
}

@article{cohen2019,
  author  = {Cohen, Aron J. and Luo, Hongjun and Guther, Kai and Dobrautz, Werner and Tew, David P. and Alavi, Ali},
  title   = {Similarity transformation of the electronic Schrödinger equation via Jastrow factorization},
  journal = {The Journal of Chemical Physics},
  volume  = {151},
  number  = {6},
  pages   = {061101},
  year    = {2019},
  month   = {08},
  issn    = {0021-9606},
  doi     = {10.1063/1.5116024},
  url     = {https://doi.org/10.1063/1.5116024}
}

@article{christlmaier2023,
  author  = {Christlmaier, Evelin Martine Corvid and Schraivogel, Thomas and López Ríos, Pablo and Alavi, Ali and Kats, Daniel},
  title   = {xTC: An efficient treatment of three-body interactions in transcorrelated methods},
  journal = {The Journal of Chemical Physics},
  volume  = {159},
  number  = {1},
  pages   = {014113},
  year    = {2023},
  month   = {07},
  issn    = {0021-9606},
  doi     = {10.1063/5.0154445},
  url     = {https://doi.org/10.1063/5.0154445}
}

@article{simula2025ecp,
  author  = {Simula, Kristoffer and Christlmaier, Evelin Martine Corvid and Filip, Maria-Andreea and Haupt, J. Philip and Kats, Daniel and Lopez-Rios, Pablo and Alavi, Ali},
  doi     = {10.1021/acs.jctc.5c00343},
  journal = {Journal of Chemical Theory and Computation},
  note    = {PMID: 40357854},
  number  = {10},
  pages   = {5155–5170},
  title   = {Transcorrelated Theory with Pseudopotentials},
  url     = { 
             
             https://doi.org/10.1021/acs.jctc.5c00343
             
             
             
             },
  volume  = {21},
  year    = {2025}
}

@article{simula2025tc_transition_metals,
  title     = {Transcorrelated theory for transition-metal atoms},
  author    = {Simula, Kristoffer and Filip, Maria-Andreea and Alavi, Ali},
  journal   = {Phys. Rev. A},
  volume    = {112},
  issue     = {3},
  pages     = {032805},
  numpages  = {12},
  year      = {2025},
  month     = {Sep},
  publisher = {American Physical Society},
  doi       = {10.1103/2ttg-789l},
  url       = {https://link.aps.org/doi/10.1103/2ttg-789l}
}

@article{haupt2023,
  author  = {Haupt, J. Philip and Hosseini, Seyed Mohammadreza and López Ríos, Pablo and Dobrautz, Werner and Cohen, Aron and Alavi, Ali},
  title   = {Optimizing Jastrow factors for the transcorrelated method},
  journal = {The Journal of Chemical Physics},
  volume  = {158},
  number  = {22},
  pages   = {224105},
  year    = {2023},
  month   = {06},
  issn    = {0021-9606},
  doi     = {10.1063/5.0147877},
  url     = {https://doi.org/10.1063/5.0147877}
}

@article{sun2020,
  author  = {Sun, Qiming and Zhang, Xing and Banerjee, Samragni and Bao, Peng and Barbry, Marc and Blunt, Nick S. and Bogdanov, Nikolay A. and Booth, George H. and Chen, Jia and Cui, Zhi-Hao and Eriksen, Janus J. and Gao, Yang and Guo, Sheng and Hermann, Jan and Hermes, Matthew R. and Koh, Kevin and Koval, Peter and Lehtola, Susi and Li, Zhendong and Liu, Junzi and Mardirossian, Narbe and McClain, James D. and Motta, Mario and Mussard, Bastien and Pham, Hung Q. and Pulkin, Artem and Purwanto, Wirawan and Robinson, Paul J. and Ronca, Enrico and Sayfutyarova, Elvira R. and Scheurer, Maximilian and Schurkus, Henry F. and Smith, James E. T. and Sun, Chong and Sun, Shi-Ning and Upadhyay, Shiv and Wagner, Lucas K. and Wang, Xiao and White, Alec and Whitfield, James Daniel and Williamson, Mark J. and Wouters, Sebastian and Yang, Jun and Yu, Jason M. and Zhu, Tianyu and Berkelbach, Timothy C. and Sharma, Sandeep and Sokolov, Alexander Yu. and Chan, Garnet Kin-Lic},
  title   = {Recent developments in the PySCF program package},
  journal = {The Journal of Chemical Physics},
  volume  = {153},
  number  = {2},
  pages   = {024109},
  year    = {2020},
  month   = {07},
  issn    = {0021-9606},
  doi     = {10.1063/5.0006074},
  url     = {https://doi.org/10.1063/5.0006074}
}

@article{needs2020,
  author  = {Needs, R. J. and Towler, M. D. and Drummond, N. D. and López Ríos, P. and Trail, J. R.},
  title   = {Variational and diffusion quantum Monte Carlo calculations with the CASINO code},
  journal = {The Journal of Chemical Physics},
  volume  = {152},
  number  = {15},
  pages   = {154106},
  year    = {2020},
  month   = {04},
  issn    = {0021-9606},
  doi     = {10.1063/1.5144288},
  url     = {https://doi.org/10.1063/1.5144288}
}

@misc{elemcojl,
  author = {D. Kats and T. Schraivogel and J. Hauskrecht and C. Rickert and F. Wu},
  title  = {{ElemCo.jl}: Julia program package for electron correlation methods},
  year   = {2024}
}

@article{Knowles89,
  author  = {Peter J. Knowles and Nicholas C. Handy},
  doi     = {10.1016/0010-4655(89)90033-7},
  issn    = {0010-4655},
  journal = {Computer Physics Communications},
  number  = {1},
  pages   = {75–83},
  title   = {A determinant based full configuration interaction program},
  url     = {https://www.sciencedirect.com/science/article/pii/0010465589900337},
  volume  = {54},
  year    = {1989}
}

@unpublished{hauskrecht2026,
  title   = {An Additive Reference Correction Scheme for the Transcorrelated Method},
  author  = {Hauskrecht, Johannes and Simula, Kristoffer and Cheng, Yifan and Christlmaier, Evelin Martine Corvid and Kats, Daniel and Alavi, Ali},
  journal = {arXiv preprint arXiv:2606.31625},
  year    = {2026}
}

@article{curtiss1991,
  author  = {Curtiss, Larry A. and Raghavachari, Krishnan and Trucks, Gary W. and Pople, John A.},
  title   = {Gaussian-2 theory for molecular energies of first- and second-row compounds},
  journal = {The Journal of Chemical Physics},
  volume  = {94},
  number  = {11},
  pages   = {7221--7230},
  year    = {1991},
  doi     = {10.1063/1.460205},
  url     = {https://doi.org/10.1063/1.460205},
  issn    = {0021-9606}
}

@article{10.1063/1.3008061,
  author  = {Feller, David and Peterson, Kirk A. and Dixon, David A.},
  title   = {A survey of factors contributing to accurate theoretical predictions of atomization energies and molecular structures},
  journal = {The Journal of Chemical Physics},
  volume  = {129},
  number  = {20},
  pages   = {204105},
  year    = {2008},
  month   = {11},
  issn    = {0021-9606},
  doi     = {10.1063/1.3008061},
  url     = {https://doi.org/10.1063/1.3008061}
}

@article{10.1063/1.478747,
  author  = {Feller, David and Peterson, Kirk A.},
  title   = {Re-examination of atomization energies for the {Gaussian-2} set of molecules},
  journal = {The Journal of Chemical Physics},
  volume  = {110},
  number  = {17},
  pages   = {8384--8396},
  year    = {1999},
  month   = {05},
  issn    = {0021-9606},
  doi     = {10.1063/1.478747},
  url     = {https://doi.org/10.1063/1.478747}
}

@article{sharma2017,
  author    = {Sharma, Sandeep and Holmes, Adam A. and Jeanmairet, Guillaume and Alavi, Ali and Umrigar, C. J.},
  title     = {Semistochastic Heat-Bath Configuration Interaction Method: Selected Configuration Interaction with Semistochastic Perturbation Theory},
  journal   = {Journal of Chemical Theory and Computation},
  volume    = {13},
  number    = {4},
  pages     = {1595--1604},
  year      = {2017},
  publisher = {American Chemical Society},
  doi       = {10.1021/acs.jctc.6b01028},
  url       = {https://doi.org/10.1021/acs.jctc.6b01028},
  issn      = {1549-9618}
}

@article{Harris2020,
  author  = {Harris, Charles R. and Millman, K. Jarrod and van der Walt, St\'efan J. and Gommers, Ralf and Virtanen, Pauli and Cournapeau, David and Wieser, Eric and Taylor, Julian and Berg, Sebastian and Smith, Nathaniel J. and Kern, Robert and Picus, Matti and Hoyer, Stephan and van Kerkwijk, Marten H. and Brett, Matthew and Haldane, Allan and del R\'io, Jaime Fern\'andez and Wiebe, Mark and Peterson, Pearu and G\'erard-Marchant, Pierre and Sheppard, Kevin and Reddy, Tyler and Weckesser, Warren and Abbasi, Hameer and Gohlke, Christoph and Oliphant, Travis E.},
  title   = {Array programming with {NumPy}},
  journal = {Nature},
  volume  = {585},
  number  = {7825},
  pages   = {357--362},
  year    = {2020},
  doi     = {10.1038/s41586-020-2649-2},
  url     = {https://doi.org/10.1038/s41586-020-2649-2},
  issn    = {1476-4687}
}

@article{bennett2017,
  author  = {Bennett, M. Chandler and Melton, Cody A. and Annaberdiyev, Abdulgani and Wang, Guangming and Shulenburger, Luke and Mitas, Lubos},
  title   = {A new generation of effective core potentials for correlated calculations},
  journal = {The Journal of Chemical Physics},
  volume  = {147},
  number  = {22},
  pages   = {224106},
  year    = {2017},
  month   = {12},
  issn    = {0021-9606},
  doi     = {10.1063/1.4995643},
  url     = {https://doi.org/10.1063/1.4995643}
}

@article{bennett2018,
  author  = {Bennett, M. Chandler and Wang, Guangming and Annaberdiyev, Abdulgani and Melton, Cody A. and Shulenburger, Luke and Mitas, Lubos},
  title   = {A new generation of effective core potentials from correlated calculations: 2nd row elements},
  journal = {The Journal of Chemical Physics},
  volume  = {149},
  number  = {10},
  pages   = {104108},
  year    = {2018},
  month   = {09},
  issn    = {0021-9606},
  doi     = {10.1063/1.5038135},
  url     = {https://doi.org/10.1063/1.5038135}
}

@incollection{karton2022,
  author    = {Karton, Amir},
  editor    = {Dixon, David A.},
  title     = {Quantum mechanical thermochemical predictions 100 years after the {S}chr\"odinger equation},
  booktitle = {Annual Reports in Computational Chemistry},
  publisher = {Elsevier},
  volume    = {18},
  chapter   = {3},
  pages     = {123--166},
  year      = {2022},
  issn      = {1574-1400},
  isbn      = {9780323990929},
  doi       = {10.1016/bs.arcc.2022.09.003},
  url       = {https://doi.org/10.1016/bs.arcc.2022.09.003}
}

@article{simula2026silicon,
  title     = {Transcorrelated wave-function framework for solids: An application to bulk and defected silicon},
  author    = {Simula, Kristoffer and Hauskrecht, Johannes and Christlmaier, Evelin Martine Corvid and Lopez-Rios, Pablo and Kats, Daniel and Usvyat, Denis and Alavi, Ali},
  journal   = {Phys. Rev. B},
  volume    = {113},
  issue     = {19},
  pages     = {195125},
  numpages  = {15},
  year      = {2026},
  month     = {May},
  publisher = {American Physical Society},
  doi       = {10.1103/d65l-5865},
  url       = {https://link.aps.org/doi/10.1103/d65l-5865}
}

@article{filip2025,
  author  = {Filip, Maria-Andreea and L\'opez R\'{\i}os, Pablo and Haupt, J. Philip and Christlmaier, Evelin Martine Corvid and Kats, Daniel and Alavi, Ali},
  title   = {Transcorrelated methods applied to second row elements},
  journal = {The Journal of Chemical Physics},
  volume  = {162},
  number  = {6},
  pages   = {064110},
  year    = {2025},
  month   = {02},
  issn    = {0021-9606},
  doi     = {10.1063/5.0246422},
  url     = {https://doi.org/10.1063/5.0246422}
}

@article{schraivogel2023,
  author  = {Schraivogel, Thomas and Christlmaier, Evelin Martine Corvid and L\'opez R\'{\i}os, Pablo and Alavi, Ali and Kats, Daniel},
  title   = {Transcorrelated coupled cluster methods. {II}. Molecular systems},
  journal = {The Journal of Chemical Physics},
  volume  = {158},
  number  = {21},
  pages   = {214106},
  year    = {2023},
  month   = {06},
  issn    = {0021-9606},
  doi     = {10.1063/5.0151412},
  url     = {https://doi.org/10.1063/5.0151412}
}

@article{filip2025jastrow,
  author  = {Filip, Maria-Andreea and Christlmaier, Evelin Martine Corvid and Haupt, J. Philip and Kats, Daniel and L\'opez R\'{\i}os, Pablo and Alavi, Ali},
  title   = {Deterministic optimization of Jastrow factors},
  journal = {The Journal of Chemical Physics},
  volume  = {163},
  number  = {8},
  pages   = {084107},
  year    = {2025},
  month   = {08},
  issn    = {0021-9606},
  doi     = {10.1063/5.0284106},
  url     = {https://doi.org/10.1063/5.0284106}
}

@article{ma2005scheme,
  title     = {Scheme for adding electron--nucleus cusps to Gaussian orbitals},
  author    = {Ma, A and Towler, MD and Drummond, ND and Needs, RJ},
  journal   = {The Journal of chemical physics},
  volume    = {122},
  number    = {22},
  year      = {2005},
  publisher = {AIP Publishing}
}

@article{boys1969determination,
  title     = {The determination of energies and wavefunctions with full electronic correlation},
  author    = {Boys, Samuel Francis and Handy, Nicholas Charles},
  journal   = {Proceedings of the Royal Society of London. A. Mathematical and Physical Sciences},
  volume    = {310},
  number    = {1500},
  pages     = {43--61},
  year      = {1969},
  publisher = {The Royal Society London}
}

@article{boys1969condition,
  title     = {A condition to remove the indeterminacy in interelectronic correlation functions},
  author    = {Boys, Samuel Francis and Handy, Nicholas Charles},
  journal   = {Proceedings of the Royal Society of London. A. Mathematical and Physical Sciences},
  volume    = {309},
  number    = {1497},
  pages     = {209--220},
  year      = {1969},
  publisher = {The Royal Society London}
}

@article{boys1969calculation,
  title     = {A calculation for the energies and wavefunctions for states of neon with full electronic correlation accuracy},
  author    = {Boys, Samuel Francis and Handy, Nicholas Charles},
  journal   = {Proceedings of the Royal Society of London. A. Mathematical and Physical Sciences},
  volume    = {310},
  number    = {1500},
  pages     = {63--78},
  year      = {1969},
  publisher = {The Royal Society London}
}

@article{luo2018combining,
  title     = {Combining the transcorrelated method with full configuration interaction quantum Monte Carlo: Application to the homogeneous electron gas},
  author    = {Luo, Hongjun and Alavi, Ali},
  journal   = {Journal of chemical theory and computation},
  volume    = {14},
  number    = {3},
  pages     = {1403--1411},
  year      = {2018},
  publisher = {ACS Publications}
}

@phdthesis{hosseini2024combining,
  title  = {Combining orbital and real space quantum Monte Carlo methods},
  author = {Hosseini, Seyed Mohammadreza},
  year   = {2024},
  school = {Dissertation, Stuttgart, Universit{\"a}t Stuttgart, 2024}
}

@article{baiardi2022explicitly,
  title     = {Explicitly correlated electronic structure calculations with transcorrelated matrix product operators},
  author    = {Baiardi, Alberto and Lesiuk, Micha{\l} and Reiher, Markus},
  journal   = {Journal of Chemical Theory and Computation},
  volume    = {18},
  number    = {7},
  pages     = {4203--4217},
  year      = {2022},
  publisher = {ACS Publications}
}

@article{corbett2025scaling,
  title     = {Scaling up the transcorrelated density matrix renormalization group},
  author    = {Corbett, Benjamin and Miyake, Akimasa},
  journal   = {Physical Review B},
  volume    = {112},
  number    = {16},
  pages     = {165120},
  year      = {2025},
  publisher = {APS}
}

@article{ammar2024compactification,
  title     = {Compactification of determinant expansions via transcorrelation},
  author    = {Ammar, Abdallah and Scemama, Anthony and Loos, Pierre-Fran{\c{c}}ois and Giner, Emmanuel},
  journal   = {The Journal of Chemical Physics},
  volume    = {161},
  number    = {8},
  year      = {2024},
  publisher = {AIP Publishing}
}

@article{haupt2026modular,
  title     = {Modular construction of Jastrow factors for the transcorrelated method},
  author    = {Haupt, J Philip and Filip, Maria-Andreea and Christlmaier, Evelin Martine Corvid and Cheng, Yifan and Hauskrecht, Johannes and Alavi, Ali},
  journal   = {The Journal of Chemical Physics},
  volume    = {164},
  number    = {11},
  year      = {2026},
  publisher = {AIP Publishing}
}

@article{sokolov2023orders,
  title     = {Orders of magnitude increased accuracy for quantum many-body problems on quantum computers via an exact transcorrelated method},
  author    = {Sokolov, Igor O and Dobrautz, Werner and Luo, Hongjun and Alavi, Ali and Tavernelli, Ivano},
  journal   = {Physical Review Research},
  volume    = {5},
  number    = {2},
  pages     = {023174},
  year      = {2023},
  publisher = {APS}
}

@article{dobrautz2024toward,
  title     = {Toward real chemical accuracy on current quantum hardware through the transcorrelated method},
  author    = {Dobrautz, Werner and Sokolov, Igor O and Liao, Ke and R{\'\i}os, Pablo L{\'o}pez and Rahm, Martin and Alavi, Ali and Tavernelli, Ivano},
  journal   = {Journal of Chemical Theory and Computation},
  volume    = {20},
  number    = {10},
  pages     = {4146--4160},
  year      = {2024},
  publisher = {ACS Publications}
}

@article{li2024variational,
  title     = {Variational quantum imaginary time evolution for matrix product state Ansatz with tests on transcorrelated Hamiltonians},
  author    = {Li, Hao-En and Li, Xiang and Huang, Jia-Cheng and Zhang, Guang-Ze and Shen, Zhu-Ping and Zhao, Chen and Li, Jun and Hu, Han-Shi},
  journal   = {The Journal of Chemical Physics},
  volume    = {161},
  number    = {14},
  year      = {2024},
  publisher = {AIP Publishing}
}

@article{Uvarov2026,
  author  = {Uvarov, Alexey and Izmaylov, Artur F.},
  title   = {Accuracy and Resource Advantages of Quantum Eigenvalue Estimation with Non-Hermitian Transcorrelated Electronic Hamiltonians},
  journal = {Journal of Chemical Theory and Computation},
  volume  = {ASAP},
  number  = {0},
  pages   = {null},
  year    = {2026},
  doi     = {10.1021/acs.jctc.6c00274},
  note    = {PMID: 42302194},
  url     = { 
             https://doi.org/10.1021/acs.jctc.6c00274
             },
  eprint  = { 
             https://doi.org/10.1021/acs.jctc.6c00274
             }
}

@article{ten2000feasible,
  title     = {A feasible transcorrelated method for treating electronic cusps using a frozen Gaussian geminal},
  author    = {Ten-no, Seiichiro},
  journal   = {Chemical Physics Letters},
  volume    = {330},
  number    = {1-2},
  pages     = {169--174},
  year      = {2000},
  publisher = {Elsevier}
}

@article{umezawa2003transcorrelated,
  title     = {Transcorrelated method for electronic systems coupled with variational Monte Carlo calculation},
  author    = {Umezawa, Naoto and Tsuneyuki, Shinji},
  journal   = {The Journal of chemical physics},
  volume    = {119},
  number    = {19},
  pages     = {10015--10031},
  year      = {2003},
  publisher = {American Institute of Physics}
}

@article{umezawa2005practical,
  title     = {A practical treatment for the three-body interactions in the transcorrelated variational Monte Carlo method: Application to atoms from lithium to neon},
  author    = {Umezawa, Naoto and Tsuneyuki, Shinji and Ohno, Takahisa and Shiraishi, Kenji and Chikyow, Toyohiro},
  journal   = {The Journal of chemical physics},
  volume    = {122},
  number    = {22},
  year      = {2005},
  publisher = {AIP Publishing}
}

@article{yanai2012canonical,
  title     = {Canonical transcorrelated theory with projected Slater-type geminals},
  author    = {Yanai, Takeshi and Shiozaki, Toru},
  journal   = {The Journal of chemical physics},
  volume    = {136},
  number    = {8},
  year      = {2012},
  publisher = {AIP Publishing}
}

@article{ten2023nonunitary,
  title     = {Nonunitary projective transcorrelation theory inspired by the F12 ansatz},
  author    = {Ten-No, Seiichiro L},
  journal   = {The Journal of Chemical Physics},
  volume    = {159},
  number    = {17},
  year      = {2023},
  publisher = {AIP Publishing}
}

@article{taube2008,
  author  = {Taube, Andrew G. and Bartlett, Rodney J.},
  title   = {Improving upon {CCSD(T)}: {$\Lambda$CCSD(T)}. {I}. Potential energy surfaces},
  journal = {The Journal of Chemical Physics},
  volume  = {128},
  number  = {4},
  pages   = {044110},
  year    = {2008},
  doi     = {10.1063/1.2830236}
}

\end{document}